\documentclass[transmag]{IEEEtran}

\usepackage[mathcal]{euscript}
\usepackage{amssymb}
\usepackage{amssymb}
\usepackage{amsmath,bm,amsthm}
\usepackage{mnsymbol}
\usepackage{graphicx}
\usepackage{psfrag}
\usepackage{epstopdf}
\usepackage{epsf,epsfig}
\usepackage{subfigure}
\usepackage{cite,color}
\usepackage{mathrsfs}
\usepackage{multirow}
\usepackage{makecell}

\usepackage{enumitem}

\usepackage{amsmath}   

\renewcommand\IEEEkeywordsname{Index Terms}
\def\BibTeX{{\rm B\kern-.05em{\sc i\kern-.025em b}\kern-.08em T\kern-.1667em\lower.7ex\hbox{E}\kern-.125emX}}
\begin{document}



%
%

\title{Joint radar and communications with multicarrier chirp-based waveform}
\author{Fredrik Berggren, \IEEEmembership{Senior Member, IEEE}, and Branislav M. Popovi\'{c}
\thanks{The authors are with Huawei Technologies Sweden AB (e-mail: \{fredrik.b,branislav.popovic\}@huawei.com).}}


\IEEEtitleabstractindextext{\begin{abstract}
We consider a multicarrier chirp-based waveform for joint radar and communication (JRC) systems and derive its time discrete periodic ambiguity function (AF). A set of waveform parameters (e.g., chirp rate) can together with the transmit sequence be selected to shape the AF to be thumbtack-like, or to be ridge-like, either along the delay axis or the Doppler axis. We demonstrate how these shapes are applicable for different use cases, e.g., radar target detection or time- and frequency synchronization. The results show that better signal detection performance than OFDM and DFT-s-OFDM can be achieved on channels with large Doppler shift. Furthermore, it is shown how transmit sequences can be selected in order to achieve low peak-to-average-power-ratio (PAPR) of the waveform.  
\end{abstract}
}
\maketitle

\section{Introduction}
\IEEEPARstart{C}{hirp} waveform with linear frequency modulation (i.e., a linear chirp) appears in various transmission systems and is, e.g., commonly used for radar \cite{Klauder}, \cite{Meta}, \cite{Rohling} and ultrasonic positioning \cite{Khyam}, as it provides precise ranging and velocity estimation. This stems from its signal properties, which enable efficient pulse-compression together with fulfilling requirements on large time-bandwidth product and low peak-to-average-power ratio (PAPR). A linear chirp has good autocorrelation properties which also makes it suitable as a synchronization signal \cite{Zhang}. Furthermore, multicarrier chirp-based waveforms have been developed for data transmission \cite{Martone},\cite{Erseghe},\cite{Ouyang},\cite{Sahin}. Chirp-convolved data transmission (CCDT) is a recently proposed multicarrier chirp waveform. This waveform includes parameters (e.g., chirp rate), which when properly selected, exhibit gains over orthogonal frequency division multiplexing (OFDM) and DFT spread OFDM (DFT-s-OFDM) in terms of lower bit- and block error rates on time-frequency selective channels with large Doppler shift \cite{Berggren}. Waveforms which perform well in such scenarios, e.g., for high speed trains, satellites \cite{38821} etc., are of interest for 5G systems \cite{He} because communications at velocities up to 350 km/h should be supported, and in some cases even as high as 500 km/h. Thereto, higher frequency bands are introduced in 5G compared to 4G systems \cite{38913}. A communication-centric joint radar and communication (JRC) system, leverages on reusing its hardware and waveform, e.g., OFDM, for radar applications \cite{Sturm}, \cite{Paul}, \cite{Liu}, \cite{Dokhanchi}. JRC systems using the chirp waveform from \cite{Ouyang} have also been suggested \cite{Lv}, \cite{Giroto}, \cite{Bhattacharjee}, showing better communication- and radar performance than for OFDM. 

A key tool for waveform synthesis is the ambiguity function (AF), which is a two-dimensional correlation function between a transmitted signal and its received time-delayed and frequency-shifted version. The AF characterizes the output of a matched filter \cite{Mercier} and is a relevant measure for analyzing and synthesizing both synchronization- and radar signals \cite{Jing},\cite{Tsao}. In this paper, we specifically consider the periodic AF, which mimics the behavior of continuous wave (CW) radar and pulse radar, cf. \cite{Freedman}, \cite{Levanon} and references therein. Periodic multicarrier radar signals, e.g., OFDM radar, could be generated by transmitting multiple OFDM symbols \cite{JWang} or through an interlaced subcarrier mapping within an OFDM symbol \cite{Mietzner}. The periodic AF is particularly convenient to study for waveforms with a cyclic prefix (CP), since the received signal undergoes a cyclic convolution with the channel impulse response. Furthermore, the time discrete AF provides valuable insight into how to design transmit sequences for radar applications \cite{Nieh}. The work in \cite{Bendetto1},\cite{Bendetto2}, \cite{Kebo} focused on constant amplitude zero autocorrelation (CAZAC) transmit sequences and analyzed the time discrete periodic AF. CAZAC sequences have many desirable properties, e.g., low PAPR due to the constant amplitude (CA) and good time-localization estimation due to zero autocorrelation (ZAC), i.e., they have an ideal autocorrelation function. For example, Zadoff-Chu (ZC) sequences are CAZAC sequences \cite{Popovic} and have been applied as reference signals, synchronization signals and random access preambles in 4G/5G systems, cf. \cite{38211}, \cite{Pitaval}. However, \cite{Bendetto1}, \cite{Bendetto2}, \cite{Kebo} only focused on the transmit sequence and did not assume any waveform, therefore the derived AFs are not directly applicable to multicarrier signals. Herein, we will close this gap and take the waveform into account when determining the AF.

Different shapes of the AF could serve different applications, e.g., a thumbtack-like AF is suitable for estimation of range and velocity for radar, or for determining the timing and the frequency offset for synchronization. With thumbtack-like, we refer to an AF that has a distinct peak at zero time- and frequency offset, while having low sidelobes otherwise. In order to obtain the time- and frequency synchronization with such a shape, the receiver could use a bank of correlators, each corresponding to a certain frequency offset hypothesis, and select the correlator output with largest magnitude \cite{Wang}. On the other hand, a ridge-like AF allows for detection of the presence of a signal under time delays or Doppler shifts. With ridge-like, we refer to an AF that has a broad peak along either the time- or frequency axis, while having low sidelobes otherwise. It has been shown that OFDM radar can decouple the range and Doppler shift, since it has an AF which is symmetric around the delay axis and around the frequency axis, respectively \cite{Franken}. In \cite{Saadia}, it was shown that DFT-s-OFDM can produce an AF with lower sidelobes than OFDM. A well-known issue with multicarrier waveforms is the high PAPR which could require substantial power back-off in the transmitter. While this issue has been studied in depth for communications, it is also recognized as important for radar \cite{Sen}, \cite{Lv2}. DFT-s-OFDM is a low-PAPR waveform which is supported in 4G/5G, cf. \cite{38211}. It has also been suggested for radar \cite{Mietzner},\cite{Saadia} and due to the DFT-precoder, the PAPR is several dB smaller than for OFDM. 

In this paper, a multicarrier chirp-based waveform, CCDT, is considered for JRC. It has been reported to outperform OFDM and DFT-s-OFDM for data transmission and would thus be a candidate waveform for JRC. However, its properties in terms of AF and PAPR were not considered in \cite{Berggren}. It has been shown that the waveforms in \cite{Ouyang},\cite{Berggren} can be represented as DFT-s-OFDM with a unitary frequency domain chirp filter. This is a big advantage since DFT-s-OFDM is already implemented in the 4G/5G terminals, and introducing CCDT could be simple and not require significant complexity increase. Furthermore, CCDT is more flexible than OFDM and DFT-s-OFDM since, as will be shown herein, with a judicious choice of the transmit sequence and parameters (e.g., the chirp rate), the AF could be shaped to be either thumbtack-like or ridge-like. 

Designing radar signals such that the AF is shaped to the environment and to a certain desired form over a given range-Doppler region, is a well-known problem, cf. \cite{Aubry}, \cite{Cui} and references therein. Flexible AF shaping is also useful for synchronization purposes, e.g., in the initial synchronization between a terminal and the base station, the AF could be ridge-like in the Doppler domain to allow the terminal to perform time-domain synchronization acquisition. Once the synchronization acquisition is achieved, the AF of the transmitted signal could be switched to thumbtack-like, to enhance time- and frequency synchronization tracking in the terminal. It should be noted that contemporary systems, e.g., 3GPP LTE and NR, do not exhibit such flexibility.

The contributions of the paper are summarized as follows.
\begin{itemize}
\item {\it Ambiguity function:} The AF is derived as a closed-form expression for an arbitrary transmit sequence and is shaped by the chirp rate and the transmit sequence. It is shown that the AF becomes the convolution in the time domain between the AF of the basis functions of the multicarrier chirp-based waveform and the AF of the transmit sequence. Moreover, we derive the AF under arbitrary sampling rate and non-integer frequency offsets.
\item {\it Transmit sequence:} We then derive the AF assuming specific transmit sequences, i.e., a ZC sequence, DFT sequence or maximum length sequence. It is shown that these can shape the AF to be either thumbtack-like or ridge-like.  
\item {\it Comparison to OFDM and DFT-s-OFDM:} We derive the AFs for OFDM and DFT-s-OFDM and compare to that of CCDT. The results show that CCDT has better detection performance on channels with large Doppler shift. 
\item{\it Joint radar and communications:}	We  show that the ambiguity functions of the communications signals based on CCDT, DFT-s-OFDM and OFDM waveforms modulated with random M-PSK symbols are of the same thumbtack type; therefore such communications signals can be used also for tracking the receiver’s range and Doppler frequency. 
\item {\it Low PAPR:} We show that ZC and DFT transmit sequences result in a signal with constant envelope.
\end{itemize}
The rest of the paper is organized as follows. In Section II, the AF of CCDT and its properties are derived. The PAPR properties are presented in Section III. Comparison to OFDM and DFT-s-OFDM is contained in Section IV. Numerical evaluation of range acquisition and range/Doppler tracking is contained in Section V, the paper is concluded in Section VI and the mathematical proofs are contained in Appendix A-F. 
 
\section{Ambiguity functions}

\subsection{Chirp-Convolved Data Transmission}
Consider the CCDT waveform for $0\le t < T$ defined by
\begin{equation}
s(t)=\sum_{m=0}^{N-1}x[m]g\left ( t - \frac{mT}{N} \right )
\label{eq:wf1}
\end{equation}
where $x[m]$, $m=0,1,\ldots,N-1$, is taken from a set of (real- or complex valued) modulation symbols. These symbols either correspond to random data or a pre-determined transmit sequence. The pulse shape (or basis function) $g(t)$ is periodic such that the time discrete representation of (\ref{eq:wf1}) for $t=nT/N$, for a symbol with samples $n=0,1,\ldots, N-1$ is defined by \cite{Berggren}
\begin{align}
s[n]&=\sum_{m=0}^{N-1} x[m]g[n-m]
\label{ccdt}\\
g[k]&=\frac{1}{\sqrt{N}}e^{-j\frac{2\pi}{N}(\alpha k^2+\beta k+ \gamma)} 
\label{basis}\\
\gcd(2\alpha,N)&=1
\label{orthogonality}\\
\alpha N +\beta  &\in \mathbb{Z}
\label{periodicity}
\end{align}
where $\alpha, \beta$ and $\gamma$ are real-valued, $\mathbb{Z}$ is the set of integers, and $\gcd(A,B)$ is the greatest common divisor of the integers $A$ and $B$. We refer to $\alpha$ as the chirp rate. A CP of length $N_{\mathrm{CP}}$ can be inserted by defining (\ref{ccdt}) for $n=-N_{\mathrm{CP}},-N_{\mathrm{CP}}+1,\ldots,-1$. The set of basis functions $g[n-m]$ are generated from cyclic time-shifts of (\ref{basis}). It has been shown that the conditions (\ref{orthogonality}) and (\ref{periodicity}) imply that (\ref{basis}) is a CAZAC sequence, i.e., the cyclically shifted basis functions are orthogonal. It has also been shown that the CCDT waveform can be represented as DFT-s-OFDM with an additional chirp filter prior to the Inverse DFT (IDFT) \cite{Berggren}. Therefore, we will also make use of the alternative representation of (\ref{ccdt})-(\ref{periodicity}) given by:
\begin{align}
s[n]&=\frac{1}{\sqrt{N}} \sum_{m=0}^{N-1} G[m]X[m]e^{j\frac{2\pi}{N}mn} \label{eq:ccdt_fd}\\
G[m]&=\sum_{k=0}^{N-1} g[k] e^{-j\frac{2\pi}{N}km} \label{eq:G}\\
X[m]&=\frac{1}{\sqrt{N}} \sum_{k=0}^{N-1} x[k] e^{-j\frac{2\pi}{N}km} \label{eq:X}
\end{align}
It can be shown that the DFT of a CAZAC sequence is a CAZAC sequence \cite{Bendetto2}. Thus $G[m]$ has CA and the filter reduces to $N$ phase shifts. Due to $G[m]$ in (\ref{eq:ccdt_fd}), the single-carrier property of DFT-s-OFDM is not maintained and modulation symbols become multiplexed in both the time- and frequency domain, thus  offering diversity gains in time-frequency selective channels. This is in contrast to OFDM where modulation symbols are frequency multiplexed, and to DFT-s-OFDM, where modulation symbols are time-multiplexed.

\subsection{Time Discrete Periodic Ambiguity Function} 
We consider the sampled low-pass equivalent signal for this analysis. For a time discrete signal $s[n]$, the periodic AF is defined as \cite{Bendetto2}
\begin{equation}
\chi (\Delta,\tau)\stackrel{\mathrm{def}}{=}\frac{1}{N}\sum_{n=0}^{N-1} s[n]s^*[n+\tau \ (\mathrm{mod}\, N)]e^{j\frac{2\pi}{N}\Delta n}
\label{peramb}
\end{equation}
for a frequency offset $\Delta \in \mathbb{Z}$ and time-delay $\tau\in \mathbb{Z}$, where $(\cdot)^*$ denotes complex-conjugate and $(\mathrm{mod}\, N)$ is the modulo-$N$ operator. In Sec. \ref{sec:nifo}, the case with non-integer frequency offsets $\Delta$ is discussed. It is straightforward to verify that (\ref{peramb}) has a period of $N$, i.e., $\chi (\Delta,\tau)=\chi (\Delta+N,\tau)=\chi (\Delta,\tau+N)=\chi (\Delta+N,\tau+N)$. The AF (\ref{peramb}) could be computed efficiently by the Inverse DFT (IDFT) for each $\tau$ of the product sequence $s[n]s^*[n+\tau \ (\mathrm{mod}\, N)]$. From Appendix A, we obtain the following main result. 

\noindent {\it Property 1. (Ambiguity function of CCDT)}  The ambiguity function is: 
\begin{align}
\chi (\Delta,\tau)&=C\sum_{m=0}^{N-1} x\left [m-\tau+\frac{r_0N-\Delta}{2\alpha} \ (\mathrm{mod} \,N)\right]x^*[m] \notag \\
&\times e^{j\frac{2\pi}{N}\Delta m} \label{eq:ambfnl}\\
C&=\sqrt{N}g^*[\tau]g[\tau+\Delta/2\alpha]g[-r_0N/2\alpha] 
e^{j\frac{2\pi}{N}\gamma}e^{j\pi \frac{r_0\Delta}{\alpha}} \\
r_0&=\frac{2\alpha k_0+\Delta +2\alpha \tau}{N} \\
2\alpha k_0 &\equiv -\Delta -2\alpha \tau \ (\mathrm{mod}\, N) \\
k_0& \in \{0,1,\ldots,N-1\}
\end{align}

Since the magnitude of $C$ is independent of $\tau$ and $\Delta$, the modulus AF of $s[n]$ is the periodic modulus AF of $x[m]$ at a delay  $\epsilon=-\tau+(r_0N-\Delta)/2\alpha$, which is a function of both $\tau$ and $\Delta$. Moreover, the  magnitude of (\ref{eq:ambfnl}) is independent of $\beta$ and $\gamma$, while the chirp rate $\alpha$ determines the delay. Thus, the AF can be shaped by the transmit sequence $x[m]$ and the parameter $\alpha$. In the following, we will only consider the modulus AF, since the phase of $C$ may not be used by the receiver. The zero Doppler cut AF is obtained by setting $\Delta=0$ in (\ref{eq:ambfnl}) and observing that $r_0=0$ is a solution to (\ref{criterion}). Hence, the modulus AF can be simplified as
\begin{align}
|\chi (\Delta=0,\tau)|=&\left | \sum_{m=0}^{N-1} x[m]x^*[m+\tau \ (\mathrm{mod} \,N)] \right |
\label{eq:pacfx}
\end{align}
which is the periodic autocorrelation function of the transmit sequence $x[m]$. Notably, it is not dependent of $\alpha$ and is thus only shaped by the transmit sequence.  

The location and magnitude of sidelobes will depend on the transmit sequence. However, a general property is that, for any $x[m]$ with CA, the AF is zero in certain locations of the $\Delta-\tau$ plane. Define the Kronecker delta function as  $\delta[k]=1$ for $k=0$ and $\delta[k]=0$ for $k\neq 0$, then from Appendix A the following property holds.

\noindent {\it Property 2. (Zeros in the $\Delta-\tau$ plane)} If $|x[m]|=1$  and $\Delta +2\alpha \tau \equiv 0 \ (\mathrm{mod} \,N)$, then  $|\chi (\Delta,\tau)|=\delta[\Delta]$.

A CAZAC sequence $x[m]$ fulfills the following CA and ZAC conditions: 
\begin{align}
|x[m]|=&1  \label{ca}\\
\frac{1}{N} \sum_{m=0}^{N-1} x[m]x^*[m+\tau \ (\mathrm{mod} \,N)]=&\delta[\tau] \label{zac}
\end{align}
It follows from Property 1 with $\Delta=0$ and (\ref{eq:pacfx}) that, if  $x[m]$ has an ideal autocorrelation function, i.e., fulfills the ZAC property (\ref{zac}), also $s[n]$ will be a ZAC sequence. However, even if $x[m]$ has CA, it does not generally guarantee that $s[n]$ has CA. Albeit, for some sequences $x[m]$ it is fulfilled, which we discuss in Sec. III. A general property for CCDT using CAZAC sequences is given from Appendix A by the following.

\noindent {\it Property 3. (Sidelobes for CAZAC)} If $x[m]$ is a CAZAC sequence, then $\sum_{\tau=0}^{N-1}|\chi (\Delta,\tau)|^2=1$.

This property can be used to determine a bound on the sidelobes of the AF. For example, if there exists a $\tau_0$ for which $|\chi (\Delta,\tau_0)|=1$, then $|\chi (\Delta,\tau_1)|=0$ for any $\tau_1 \neq \tau_0$.

\begin{figure*}
\begin{minipage}{.33\textwidth}
\begin{center}
\includegraphics[width=\textwidth, height=60 mm]{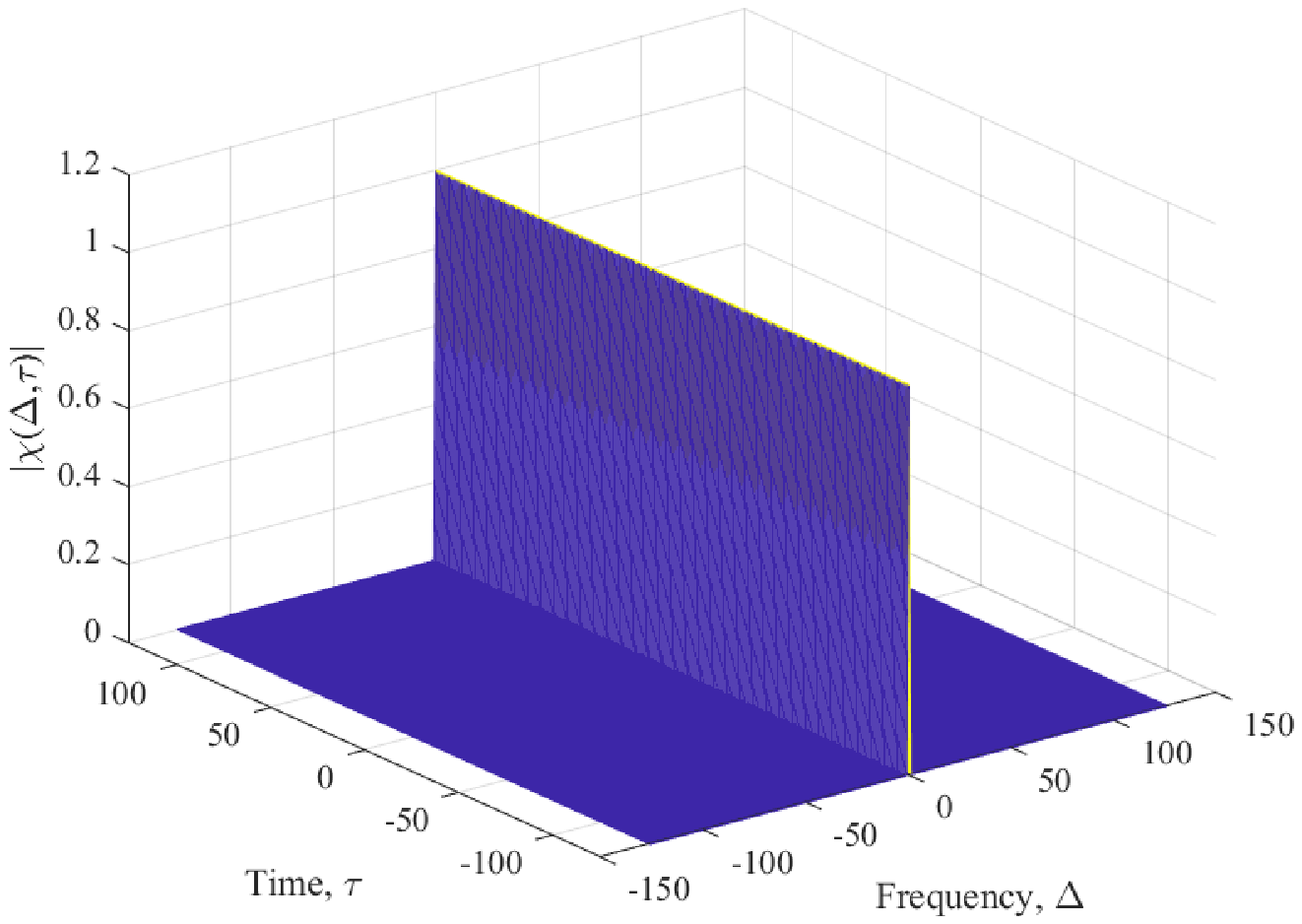}
\text{(a)}
\end{center}
\end{minipage}%
\begin{minipage}{.33\textwidth}
\begin{center}
\includegraphics[width=\textwidth, height=60 mm]{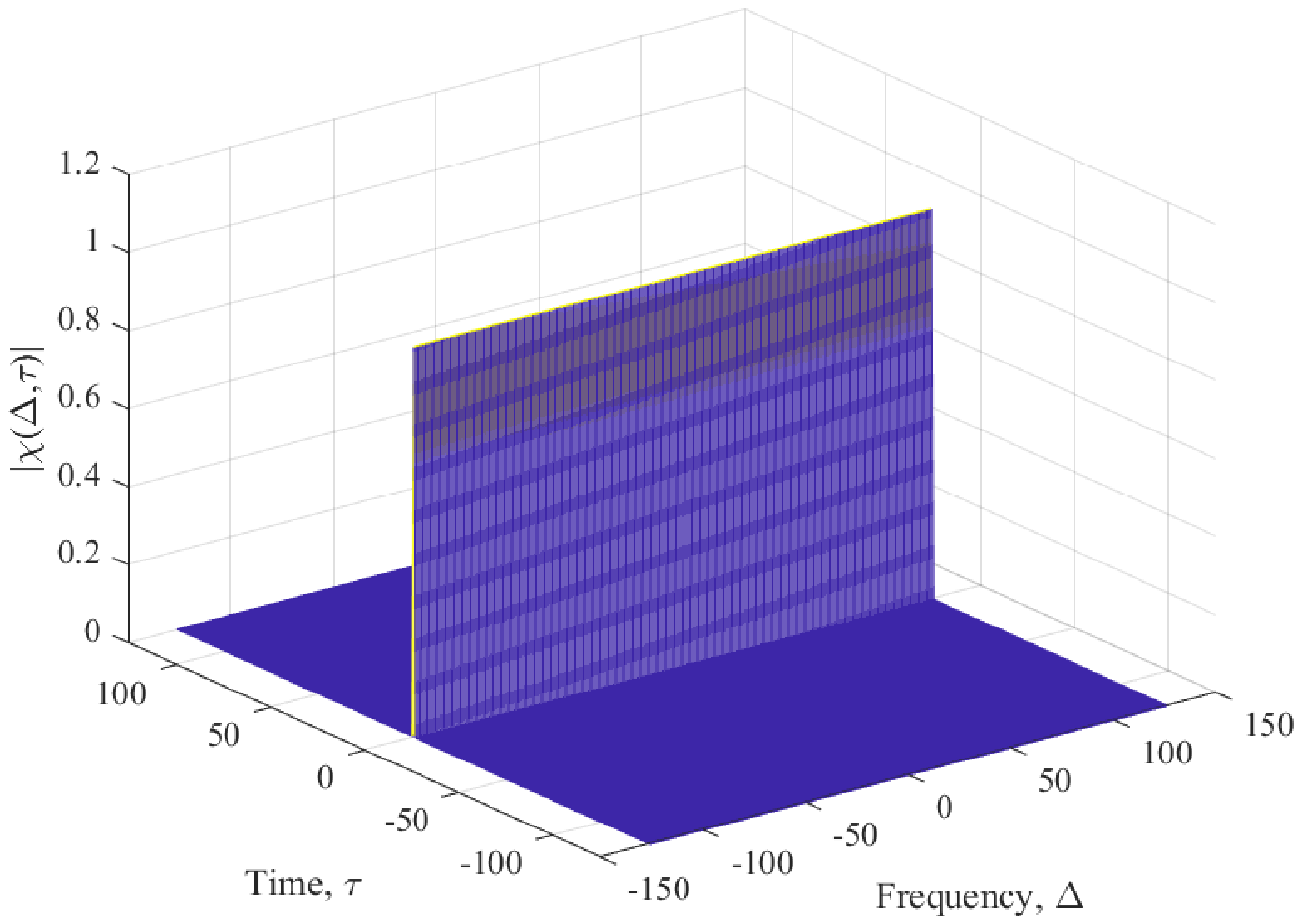}
\text{(b)}
\end{center}
\end{minipage}%
\begin{minipage}{.33\textwidth}
\begin{center}
\includegraphics[width=\textwidth, height=60 mm]{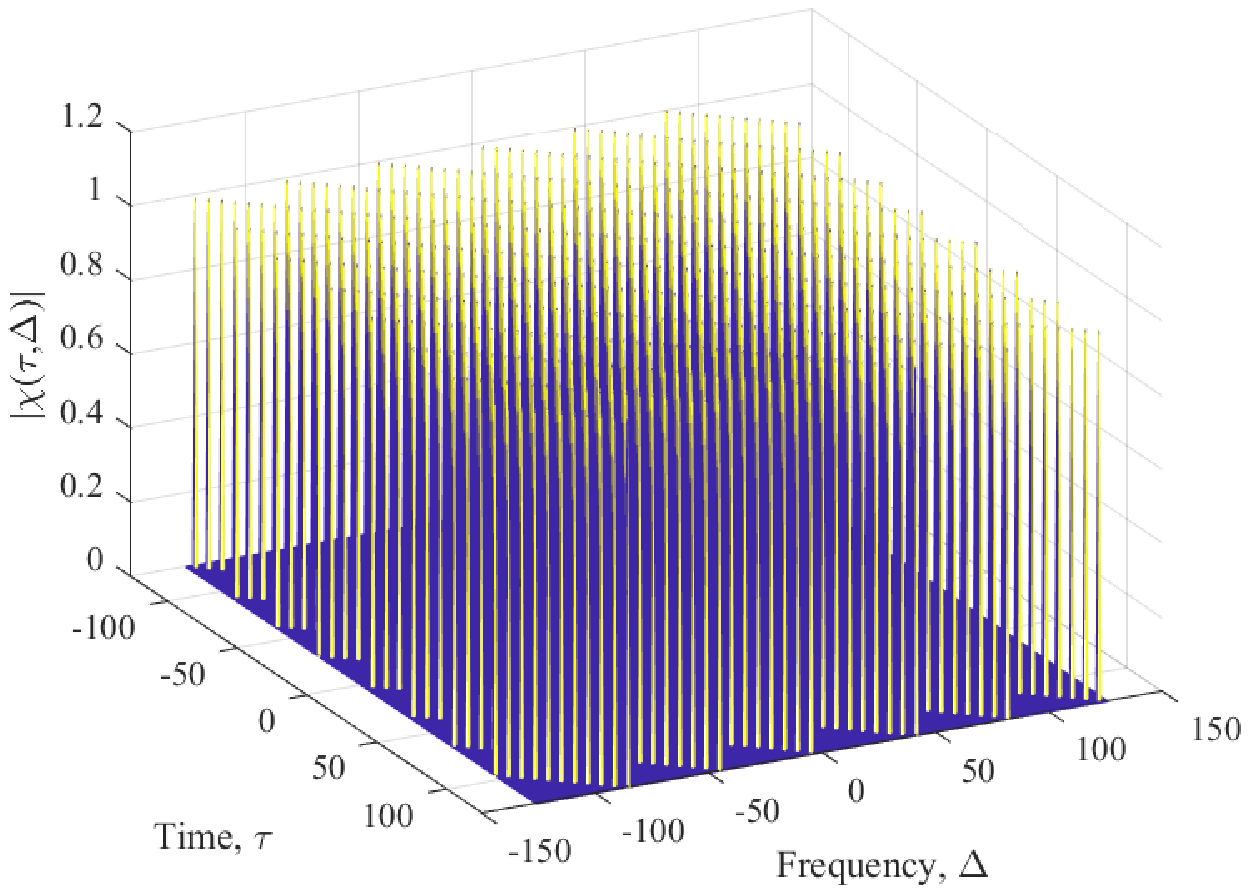}
\text{(c)}
\end{center}
\end{minipage}%
\caption{Modulus AF based on different transmit sequences of length $N=127$. (a) A ridge along the time domain from a DFT sequence. (b) A ridge along the frequency domain from a Zadoff-Chu sequence with $\alpha=u/2$. (c) An AF without ridge from a Zadoff-Chu sequence with $\alpha=2u$.}
\label{fig:ridges}
\end{figure*}

A ridge-like AF makes it possible to detect the presence of a signal by a matched filter under any time delay or frequency shift.
According to Property 4 shown in Appendix B, a ridge-like AF is generated from a DFT sequence, which is illustrated in Fig. \ref{fig:ridges} (a). The ridge is along the $\tau$-axis, i.e., $\tau$ and $\Delta$ are decoupled.  

\noindent {\it Property 4. (AF for DFT-sequence)} If $x_k[m]=e^{j\frac{2\pi}{N}km}$ for $k=0,1,\ldots,N-1$, then $|\chi (\Delta,\tau)|=\delta[\Delta]$.

In this case, the ridge lies along the time axis and thus the detection of the presence of a signal with any time delay could be performed by a matched filter, which will produce the maximum value for any time delay. This type of AF could also be applicable for estimating the frequency offset, e.g., by fixing a time delay and performing matched filtering with one filter for each frequency offset hypothesis. Thereby, the frequency offset is determined from the hypothesis that produces the largest matched filter output.

An advantage of the CCDT waveforms is the possibility to shape their AFs by modifying their parameter $\alpha$. According to Property 5 below (proven in Appendix B), the AF can be shaped such that
the ridge is rotated 90 degrees in the $\Delta-\tau$ plane, as illustrated in Fig. \ref{fig:ridges} (b).  

\noindent {\it Property 5. (AF for ZC sequence)} 
If $x_u[m]=e^{j\frac{\pi}{N}um(m+1)}$, where $N$ is odd, or $x_u[m]=e^{j\frac{\pi}{N}um^2}$, where $N$ is even, and $\alpha=u/2$, then $|\chi (\Delta,\tau)|=\delta[\tau]$

In this case, the ridge lies along the frequency axis and thus the detection of the presence of a signal with any Doppler shift could be performed by a matched filter, which will produce the maximum value for any Doppler shift. A ridge-like AF of the CCDT waveform can be generated from a ZC sequence with arbitrary root index $u$. Other selections of $\alpha$ will generate different shapes, as illustrated by Fig. \ref{fig:ridges} (c).

%

\subsection{Thumbtack-like Ambiguity Function}
While ridge-like AFs can detect the presence of a signal, a thumbtack-like AF may be needed for estimating parameters such as range, velocity and synchronization timing. A thumbtack-like AF is characterized by having a small value of $| \chi (\Delta,\tau)|$ for $\Delta\neq 0$ and $\tau \neq 0$. This allows unambiguous time- and frequency synchronization for communications or range and velocity estimation for radar. An interesting case is where a maximum length sequence, aka. m-sequence, is used. The m-sequence is defined for $N=2^p-1$ for a positive integer $p$, $x[m]\in \{-1,1\}$ and its periodic autocorrelation function is
\begin{align}
\rho(\tau)=\frac{1}{N}\sum_{m=0}^{N-1} x[m]x[m+\tau\ (\mathrm{mod} \,N)]=
\begin{cases}
1, & \tau=0\\
-\frac{1}{N}, & \tau \neq 0.
\end{cases}
\label{corrmseq}
\end{align}
The modulus AF is given by Property 6 shown in Appendix B, which is illustrated in Fig. \ref{fig:thumbs}.

\noindent {\it Property 6. (AF of m-sequence)} If $x[m]$ is an m-sequence, then
\begin{equation}
| \chi (\Delta,\tau)|=
\begin{cases}
1, & \Delta=0, \tau=0\\
\frac{1}{N}, & \Delta=0, \tau \neq 0\\
0, & \Delta +2\alpha \tau \equiv 0 \ (\mathrm{mod} \,N)\\
 \sqrt{(N+1)}/N, & \mathrm{otherwise.}
\end{cases}
\label{mseq}
\end{equation}


\begin{figure*}
\begin{minipage}{.5\textwidth}
\begin{center}
\includegraphics[width=\textwidth, height=70 mm]{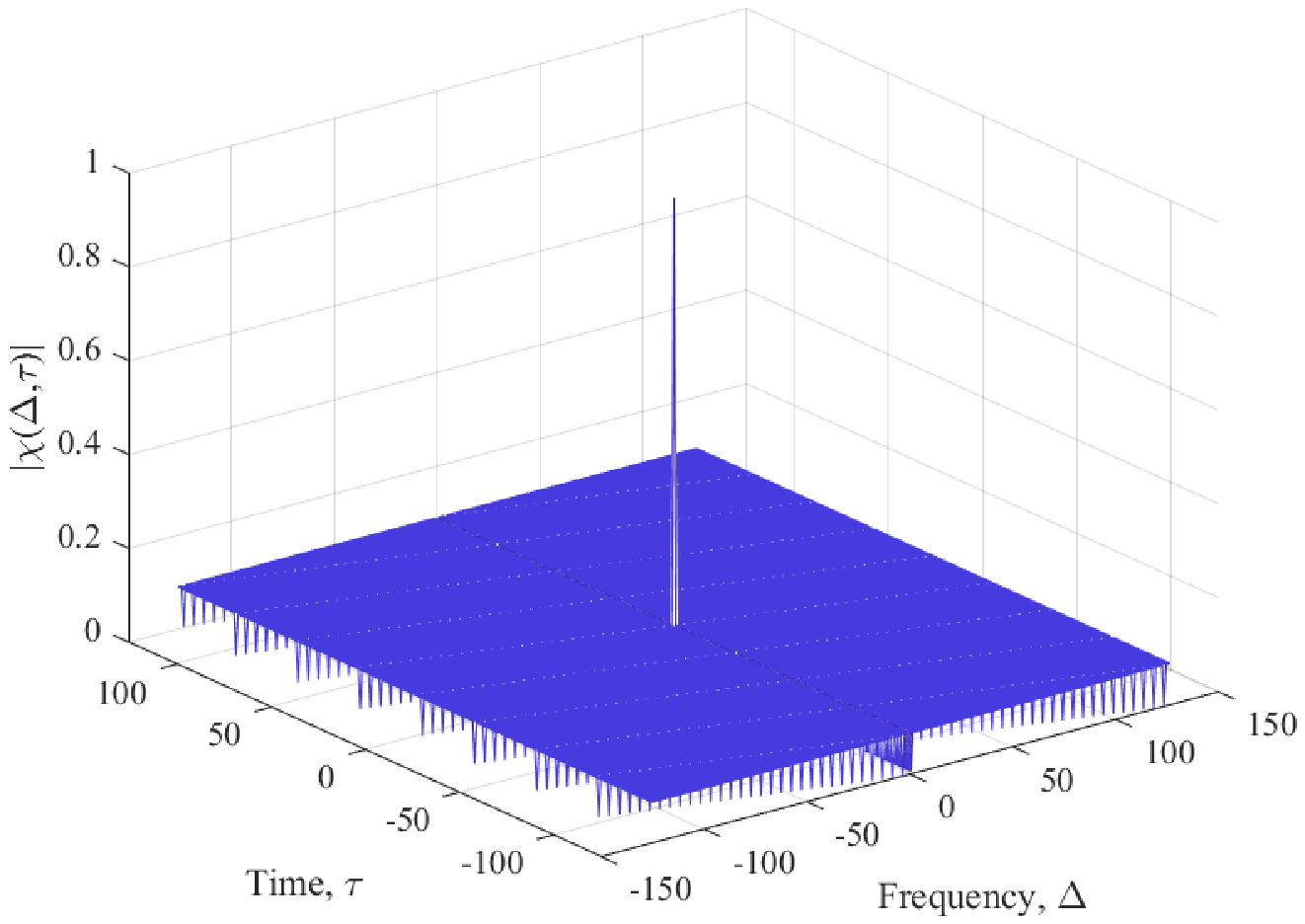}
\text{(a)}
\end{center}
\end{minipage}%
\begin{minipage}{.5\textwidth}
\begin{center}
\includegraphics[width=\textwidth, height=70 mm]{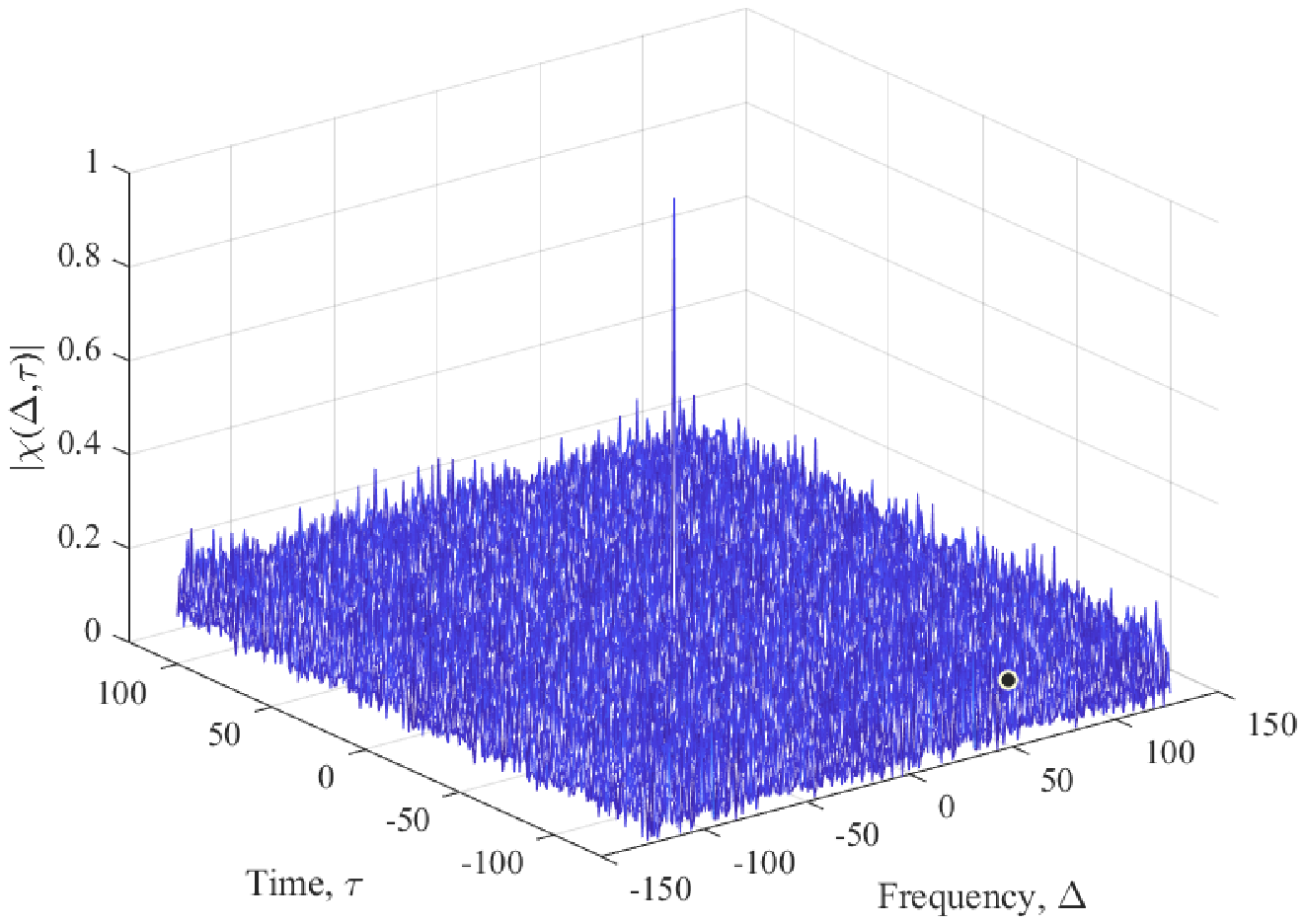}
\text{(b)}
\end{center}
\end{minipage}%
\caption{Modulus AF based on different transmit sequences of length $N=127$ with $\alpha=2$ and $\beta=1$. (a) m-sequence, (b) random unitary modulation symbols.}
\label{fig:thumbs}
\end{figure*}

\subsection{Ambiguity function with random data}
\label{Sec:random}
For JRC systems, $x[m]$ may consist of random, but for the transmitter known, modulation symbols, e.g., used for transmitting data. It is therefore important that the expected AF has good properties.  Analysis of the statistical properties of the AF has similarly been performed for noise radar systems, where the transmitted signal is obtained from a stochastic process \cite{Pralon}. 
Suppose $x[m]=e^{-j2\pi\phi_m'}$ and $\phi_m'$ is chosen independently and randomly from a uniform distribution $\phi_m'\in[0,1)$. Let $\epsilon=-\tau+(r_0N-\Delta)/2\alpha\neq 0$, then it follows that for $\Delta  \neq 0$ and $\tau\neq 0$, $x[m]x^*[m+\epsilon \ (\mathrm{mod} \,N)]=e^{j2\pi(\phi_{m+\epsilon \; (\mathrm{mod} \,N)}'-\phi_m')}$ and $\phi_m=\phi_{m+\epsilon \; (\mathrm{mod} \,N)}'-\phi_m'$ becomes a random variable. The probability density function of the difference between two uniform random variables can be determined as
\begin{equation}
f_{\phi_m}(x)=
\begin{cases}
x+1, & -1<x\le 0\\
1-x, & 0<x<1.
\end{cases}
\label{eq:pdf}
\end{equation}

Utilizing $\int xe^{ax} dx=(ax-1)a^{-2}e^{ax}$, we can obtain the expectation value, $\mathbb{E}[\cdot]$, of (\ref{eq:ambfnl}) as

\begin{align}
| \mathbb{E}\left [ \chi (\Delta  \neq 0,\tau\neq 0) \right ] |=& \frac{1}{N} \left |  \sum_{m=0}^{N-1}\mathbb{E}\left [ e^{j2\pi\phi_m}\right ]e^{j\frac{2\pi}{N}\Delta m} \right | \notag\\
=& \frac{1}{N} \Bigg | \sum_{m=0}^{N-1} \left ( \int_{-1}^{1} f_{\phi_m}(\phi)e^{j2\pi\phi} d\phi \right ) \notag \\
\times &e^{j\frac{2\pi}{N}\Delta m} \Bigg  | \notag \\
=&0.
\label{eq:rndcnt}
\end{align}

Thus the expectation value of the modulus AF exhibits a thumbtack-like shape and good detection performance in average sense is expected. Fig. \ref{fig:thumbs} shows one realization of the modulus AF where the modulation symbols are randomly generated on the unit circle.  Moreover, if the modulation symbols are chosen independently and randomly from an $M$-PSK constellation, $\phi_m'=e^{j\frac{2\pi}{M}p}$ with $\mathrm{Pr}[p=p']=1/M$ and $p=0,1,\ldots, M-1$, then $\phi_m$ will correspond to the angle of one of the constellation points with uniform probability. Therefore, the expectation value of the AF will be thumbtack-like since, by using (\ref{geosum})
\begin{equation}
\begin{aligned}
|\mathbb{E}\left [ \chi (\Delta \neq 0,\tau\neq 0) \right ]|=&\frac{1}{N} \left | \sum_{m=0}^{N-1}\mathbb{E}\left [ e^{j2\pi\phi_m}\right ]e^{j\frac{2\pi}{N}\Delta m} \right |\\
=&\frac{1}{N} \left |  \sum_{m=0}^{N-1} \left ( \sum_{p=0}^{M-1} \frac{1}{M} e^{j\frac{2\pi}{M}p}\right ) e^{j\frac{2\pi}{N}\Delta m} \right |\\
=&0.\label{eq:rndpsk}
\end{aligned}
\end{equation}
Thus CCDT would be suitable for JRC, where the $x[m]$ represents modulation symbols, which are known but not pre-determined.

\subsection{Ambiguity Function With Upsampling}

The AF of an upsampled signal can be obtained from (\ref{eq:ccdt_fd}) by replacing $N$ with  $Q \, (Q>N)$ in the exponential function. As shown in Appendix C, when $Q/N \in \mathbb{Z}$, the AF becomes 
\begin{align}
\chi (\Delta,\tau)=& \frac{N}{Q} \sum_{v=0}^{N-1} \sum_{w=0}^{N-1}\chi_g (\Delta,v)  \chi_x (\Delta,w)\notag \\
 \times &\frac{\sin \left (\pi \left (v+w-\frac{N}{Q}\tau \right ) \right ) }{\sin \left ( \frac{\pi \left (v+w-\frac{N}{Q}\tau \right ) }{N}\right )}
e^{\frac{\pi(N-1)\left (v+w-\frac{N}{Q}\tau \right )}{N}}
\label{eq:ambu2}
\end{align}
where $\chi_g (\Delta,\tau)$ and $\chi_x (\Delta,\tau)$ are the AFs of the sequences $g[k]$ and $x[m]$, respectively.
Fig. \ref{fig:thumbs2} shows the modulus AF with $Q=10N$, using an m-sequence for $x[m]$. This should be compared to Fig. \ref{fig:thumbs} (i.e., where $Q=N$) and it can be seen that the thumbtack-like characteristics are maintained with upsampling. An interesting case of (\ref{eq:ambu2}) is when $Q=N$, which gives an alternative expression for the AF with no upsampling.  From Appendix C, we obtain
\begin{align}
\chi (\Delta,\tau)=&N \sum_{v=0}^{N-1} \chi_g (\Delta,v) \chi_x (\Delta,\tau-v). 
\label{eq:conva}
\end{align}
In other words, the convolution over the delays of the AFs $\chi_g (\Delta,\tau)$ and $\chi_x (\Delta,\tau)$ gives the AF for $s[n]$. The AF of (\ref{eq:ambfnl}) could equivalently be obtained by inserting $\chi_g (\Delta,\tau)$ and $\chi_x (\Delta,\tau)$ in (\ref{eq:conva}), which is shown in Appendix D. 

\subsection{Ambiguity Function With Non-integer Frequency Offset}
\label{sec:nifo}

\begin{figure*}
\begin{minipage}{.5\textwidth}
\begin{center}
\includegraphics[width=\textwidth, height=70 mm]{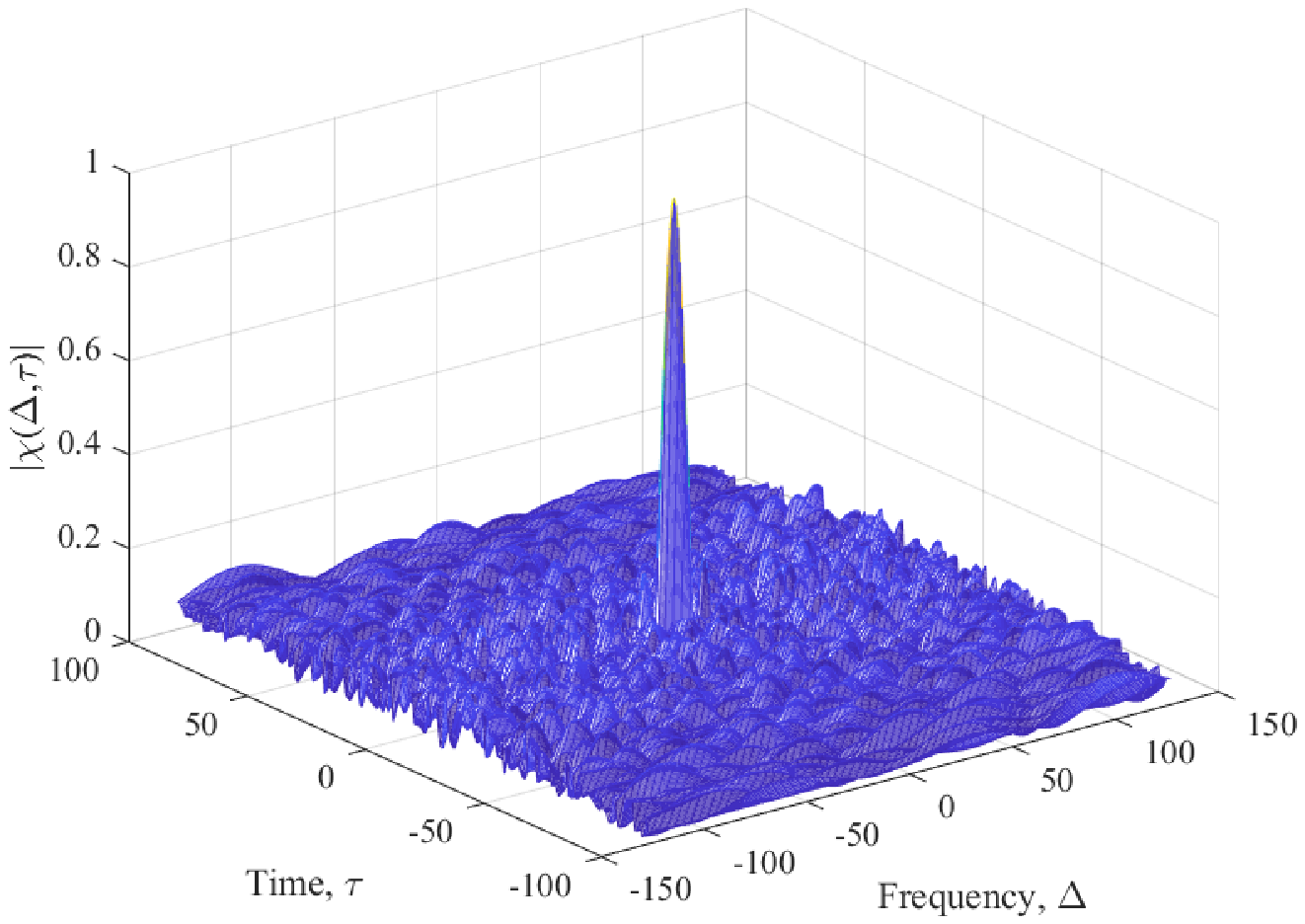}
\text{(a)}
\end{center}
\end{minipage}%
\begin{minipage}{.5\textwidth}
\begin{center}
\includegraphics[width=\textwidth, height=70 mm]{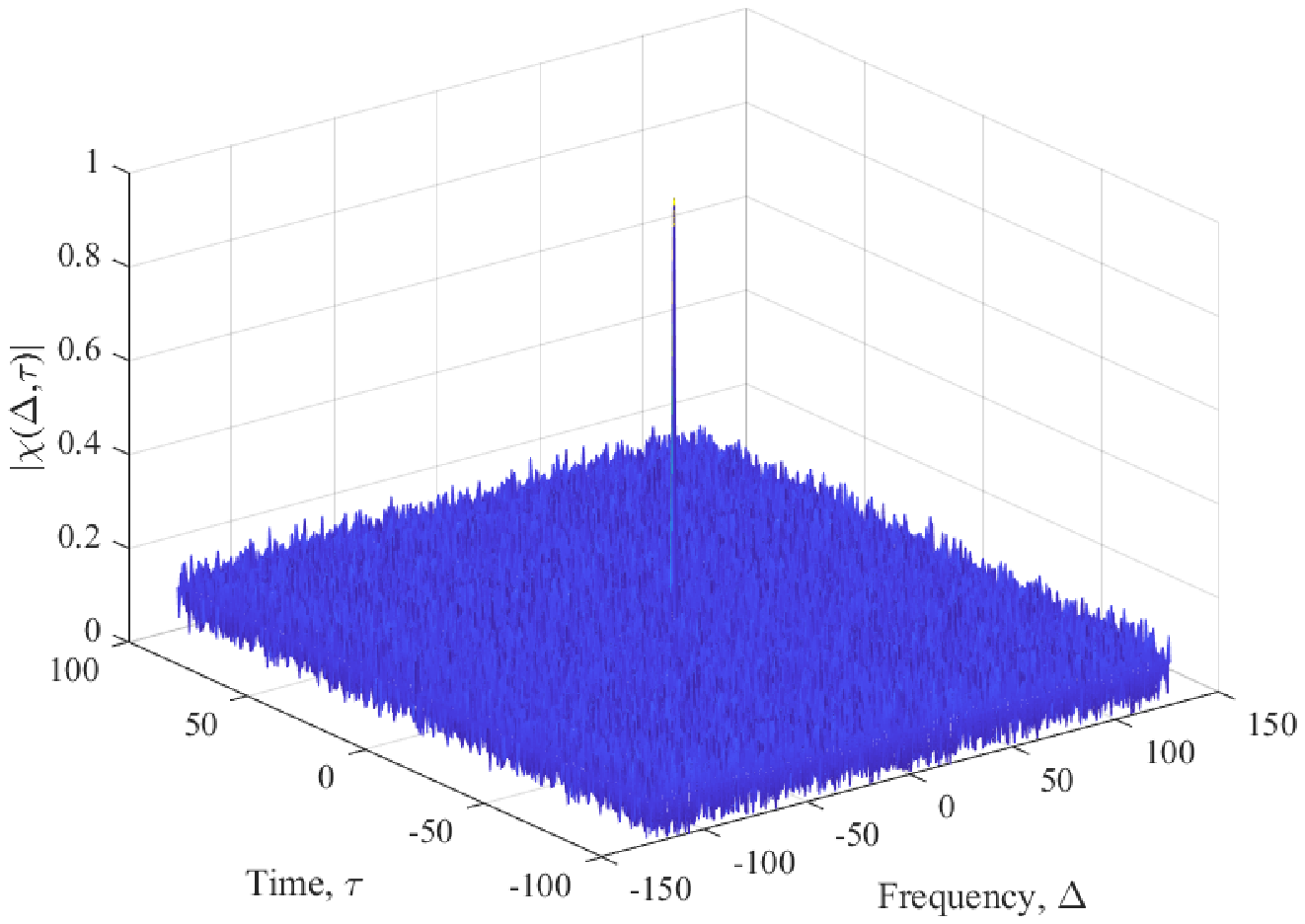}
\text{(b)}
\end{center}
\end{minipage}%
\caption{Modulus AF based based on an m-sequence of length $N=127$ with $\alpha=2$ and $\beta=1$. (a) with $Q/N=10$ times upsampling, (b) with non-integer frequency offset.}
\label{fig:thumbs2}
\end{figure*}

The AF with upsampling and non-integer frequency offset are given by (\ref{eq:upgen}) and (\ref{eq:upgen2}) in Appendix C. Further simplification can be done for the case without upsampling by using this identity for an arbitrary $p$
\begin{align}
\sum_{n=0}^{N-1} e^{j\frac{2\pi}{N}np}&=\frac{\sin (\pi p)}{\sin \left (\frac{\pi p}{N}\right )}e^{j\frac{\pi (N-1)p}{N}} \label{eq:geosum2}
\end{align}
in (\ref{amb1}) with $p=\Delta - 2\alpha (m-k)+2\alpha \tau$. Then the AF becomes as follows.
\begin{align}
\chi (\Delta,\tau)&= \sum_{m=0}^{N-1}\sum_{k=0}^{N-1} x[k]x^*[m] e^{-j\frac{2\pi}{N}\left(\alpha\left(k^2-m^2\right ) +\beta (m-k)+2\alpha\tau m\right )} \notag \\
& \times C_0  \frac{\sin (\pi (\Delta - 2\alpha (m-k)+2\alpha \tau))}{\sin \left (\frac{\pi (\Delta - 2\alpha (m-k)+2\alpha \tau)}{N}\right )} \notag \\
& \times e^{j\frac{\pi (N-1)(\Delta - 2\alpha (m-k)+2\alpha \tau)}{N}}
\label{ambni}
\end{align}
Fig. \ref{fig:thumbs2} shows the modulus AF (\ref{ambni}) when $\Delta$ is assumed in steps of 0.1, and similarly to Fig. \ref{fig:thumbs}, the thumbtack-like shape is maintained.

\section{Peak-to-Average-Power-Ratio}
A low PAPR allows less power back-off in the transmitter and thus has benefits for coverage of the transmitted signal.
If $|x[m]|=1$, it can be shown using (\ref{ccdt}) that $\frac{1}{N} \sum_{n=0}^{N-1} |s[n]|^2=1$. Thus, the PAPR (without upsampling) is defined as:
\begin{equation}
\mathrm{PAPR}=10 \log_{10} \left ( \max_{0\le n \le N-1} |s[n]|^2 \right )
\end{equation}
A CA sequence has by definition 0 dB PAPR, but that does not generally guarantee that $s[n]$ has 0 dB PAPR. However, the following properties for ZC and DFT sequences shown in Appendix E, confirm that it is the case for CCDT.

\noindent {\it Property 7. (PAPR using ZC sequence)} If $x_u[m]=e^{j\frac{\pi}{N}um(m+1)}$, where $N$ is odd prime and $\mathrm{gcd}(u,N)=1$, 
\begin{equation*}
\mathrm{PAPR}=
\begin{cases}
10 \log_{10} N, & u=2\alpha\\
0, & u\neq 2\alpha.
\end{cases}
\end{equation*}
From (\ref{eq:proof13}), it follows that when $u=2\alpha$ there exists an $n=n_0$ fulfilling $\alpha+2\alpha n_0+\beta \equiv 0 \ (\mathrm{mod} \,N)$ and the signal becomes
\begin{equation}
s[n]=
\begin{cases}
\sqrt{N}e^{-j\frac{2\pi}{N}(\alpha n_0^2+\beta n_0 +\gamma)}, & n=n_0\\
0, & n\neq n_0.
\end{cases}
\end{equation}
Thus, the PAPR of $10 \log_{10} N$ for $u=2\alpha$ is a consequence of that the signal contains all energy in one sample, $n_0$. 
For any DFT sequence, $\mathrm{PAPR}=0$ dB, as shown by the following property.

\noindent {\it Property 8. (PAPR using DFT sequence)} If $x_k[m]=e^{j\frac{2\pi}{N}km}$, for $k=0,1,\ldots,N-1$, $\mathrm{PAPR}=0$ dB.

\section{Comparison of AF and PAPR with OFDM and DFT-s-OFDM}
\subsection{Ambiguity Functions}
As mentioned in Sec. II.A, CCDT is utilizing features of both OFDM (i.e., transmitting a modulation symbol over the whole OFDM symbol) and DFT-s-OFDM (i.e., transmitting a modulation symbol over the whole bandwidth). Therefore, we will compare the AF of CCDT with those of OFDM and DFT-s-OFDM. Such expressions appear not to be available in the literature, and for completeness we derive in Appendix F the AFs with different transmit sequences. 
The OFDM signal is defined by
\begin{equation}
s[n]=\frac{1}{\sqrt{N}} \sum_{k=0}^{N-1} x[k]e^{j\frac{2\pi }{N}kn}
\end{equation}
which when inserted in (\ref{peramb}) yields
\begin{align}
|\chi_\mathrm{OFDM} (\Delta,\tau)|&=\frac{1}{N^2}\Bigg | \sum_{m=0}^{N-1} \sum_{k=0}^{N-1}x[m]x^*[k]e^{-j\frac{2\pi}{N}\tau k} \notag\\
&\times \sum_{n=0}^{N-1} e^{j\frac{2\pi }{N}n(m-k+\Delta)} \Bigg | \notag \\
&=\!\! \frac{1}{N} \left | \sum_{m=0}^{N-1}x[m-\Delta \ (\mathrm{mod} \,N)]x^*[m] e^{-j\frac{2\pi}{N}\tau m} \right |
\label{ofdmamb}
\end{align}
since the inner sum is equal to $N\delta [m-k+\Delta  \ (\mathrm{mod} \,N)]$ due to (\ref{geosum}). The DFT-s-OFDM signal is defined by inserting a DFT-precoder prior to the OFDM modulator such that
\begin{align}
s[n]&=\frac{1}{\sqrt{N}} \sum_{k=0}^{N-1} \frac{1}{\sqrt{N}}\sum_{l=0}^{N-1} x[l]e^{-j\frac{2\pi kl}{N}}e^{j\frac{2\pi kn}{N}} \notag \\
&=x[n]
\end{align}
which when inserted in (\ref{peramb}) yields
\begin{align}
|\chi_\mathrm{DFT-s-OFDM} (\Delta,\tau)|&=\frac{1}{N} \Big | \sum_{m=0}^{N-1} x[m-\tau \ (\mathrm{mod} \,N)]x^*[m] \notag \\
&\times e^{j\frac{2\pi}{N}\Delta m} \Big |.
\label{dftsofdmamb}
\end{align}
There are several differences between (\ref{eq:ambfnl}) and (\ref{ofdmamb}), e.g., it is the periodic AF of $x[m]$ but with $\Delta$ and $\tau$ interchanged compared to  (\ref{eq:ambfnl}). The similarity between (\ref{eq:ambfnl}) and (\ref{dftsofdmamb}) is that (\ref{dftsofdmamb}) is also the periodic ambiguity autocorrelation function of $x[m]$, but at a delay, $\tau$, which is independent of $\Delta$. 

Further simplifications can be done, e.g., the OFDM zero delay cut AF follows directly from (\ref{ofdmamb}) as
\begin{align}
|\chi_\mathrm{OFDM} (\Delta,\tau=0)|&=\frac{1}{N} \left | \sum_{m=0}^{N-1}x[m-\Delta \ (\mathrm{mod} \,N)]x^*[m] \right |
\label{ofdmpacf}
\end{align}
and is thus the periodic autocorrelation function of $x[m]$. The DFT-s-OFDM zero Doppler cut AF follows directly from (\ref{dftsofdmamb}) as
\begin{align}
|\chi_\mathrm{DFT-s-OFDM} (\Delta=0,\tau)|&=\frac{1}{N} \Big | \sum_{m=0}^{N-1} x[m-\tau \ (\mathrm{mod} \,N)] \notag \\
&\times x^*[m] \Big |
\end{align}
and is thus the periodic autocorrelation function of $x[m]$. Furthermore, the OFDM zero Doppler cut AF is an impulse as shown by the following.

\noindent {\it Property 9. (Zero Doppler cut AF for OFDM)} When $|x[m]|=1$,  then $|\chi_\mathrm{OFDM} (\Delta=0,\tau)|=\delta[\tau]$.

The property implies that an OFDM signal with constant modulus modulation symbols has ideal periodic autocorrelation, which was also shown in \cite{Popovic2}. For DFT-s-OFDM, a related property can be derived.

\noindent {\it Property 10. (Zero delay cut AF for DFT-s-OFDM)} When $|x[m]|=1$,  then $|\chi_\mathrm{DFT-s-OFDM} (\Delta,\tau=0)|=\delta[\Delta]$


A ridge-like shape could be achieved from a DFT sequence according to the following property. It should be noted that this ridge is the same as for CCDT with ZC sequence (i.e., Property 5 and a ridge along the frequency axis Fig. \ref{fig:ridges} (b)).

\noindent {\it Property 11. (AF for DFT-sequence for OFDM)} If $x_k[m]=e^{j\frac{2\pi}{N}km}$, for $k=0,1,\ldots,N-1$, then $|\chi_\mathrm{OFDM} (\Delta,\tau)|=\delta[\tau]$.

Moreover, a ridge-like AF can be produced, similarly as for CCDT with a DFT sequence (i.e., Property 4 and a ridge  along the time axis Fig. \ref{fig:ridges} (a)).

\noindent {\it Property 12. (AF for DFT sequence for DFT-s-OFDM)} If $x_k[m]=e^{j\frac{2\pi}{N}km}$, for $k=0,1,\ldots,N-1$, then $|\chi_\mathrm{DFT-s-OFDM} (\Delta,\tau)|=\delta[\Delta]$.

In contrast to CCDT, a ZC sequence does not produce an AF with a ridge with decoupled $\tau$ and $\Delta$ parameters, which is shown by the following property.

\noindent {\it Property 13. (AF for ZC sequence for OFDM)} If $x_u[m]=e^{j\frac{\pi}{N}um(m+1)}$, $N$ is odd and $u=1,\ldots,N-1$, then $|\chi_\mathrm{OFDM} (\Delta,\tau)|=\delta[u\Delta +\tau \ (\mathrm{mod} \,N)] $.

Notably, a ZC sequence was used for the primary synchronization signal (PSS) in 3GPP LTE. However, it was replaced by an m-sequence in 3GPP NR, much due to the undesirable sidelobles and the coupling of $\tau$ and $\Delta$ in the AF. As given by Property 13 for OFDM, a ZC sequence does not produce an AF with a ridge with decoupled $\tau$ and $\Delta$ parameters. This is also the case for DFT-s-OFDM, which is shown by the following property.

\noindent {\it Property 14. (AF for ZC sequence for DFT-s-OFDM)} If $x_u[m]=e^{j\frac{\pi}{N}um(m+1)}$, $N$ is odd and $u=1,\ldots,N-1$, then $|\chi_\mathrm{DFT-s-OFDM} (\Delta,\tau)|=\delta[\Delta-\tau u \ (\mathrm{mod} \,N)]$.

The AFs are summarized in Table \ref{table_amb}, showing that the modulation sequences can be chosen such that CCDT produces a ridge  along either the time- or frequency axis, whereas for OFDM the ridge is along the time axis and for DFT-s-OFDM the ridge is along the frequency axis. For ZC sequences, neither OFDM or DFT-s-OFDM exhibit an AF with decoupled $\tau$ and $\Delta$, as for CCDT. Furthermore, by using (\ref{ofdmamb}) or (\ref{dftsofdmamb}), it will be possible to use the same steps as for (\ref{eq:pdf})-(\ref{eq:rndpsk}) and show that when random unitary modulation symbols are used, the expected AF is thumbtack-like. Hence, the whole family of CCDT, OFDM and DFT-s-OFDM waveforms may be suited for JRC systems. 

\begin{table*}
\caption{Comparison of ambiguity functions.}
\label{table_amb}
\begin{center}
\begin{tabular}{c|c|c|c}
\hline
\hline
Sequence & OFDM & DFT-s-OFDM & CCDT\\
\hline
DFT & $|\chi_\mathrm{OFDM} (\Delta,\tau)|=\delta[\tau]$ & $|\chi_\mathrm{DFT-s-OFDM} (\Delta,\tau)|=\delta[\Delta]$ & $|\chi (\Delta,\tau)|=\delta[\Delta]$\\
\hline
ZC & $|\chi_\mathrm{OFDM} (\Delta,\tau)|=\delta[u\Delta +\tau \ (\mathrm{mod} \,N)] $ &$|\chi_\mathrm{DFT-s-OFDM} (\Delta,\tau)|=\delta[\Delta-\tau u \ (\mathrm{mod} \,N)]$ &\multirow[t]{2}{*}{$|\chi (\Delta,\tau)|=\delta[\tau]$}\\ & & & if $u=2\alpha$ \\
\hline
$|x[m]|=1$ & $|\chi_\mathrm{OFDM} (\Delta=0,\tau)|=\delta[\tau]$ & $|\chi_\mathrm{DFT-s-OFDM} (\Delta,\tau=0)|=\delta[\Delta]$ & \multirow[t]{2}{*}{$|\chi (\Delta,\tau)|=\delta[\Delta]$ } \\ & & & if $\Delta +2\alpha \tau \equiv 0 \ (\mathrm{mod} \,N)$ \\
\hline 
\hline
\end{tabular}
\end{center}
\end{table*}

\subsection{PAPR}
\begin{figure}
\begin{center}
\includegraphics[width=0.50\textwidth]{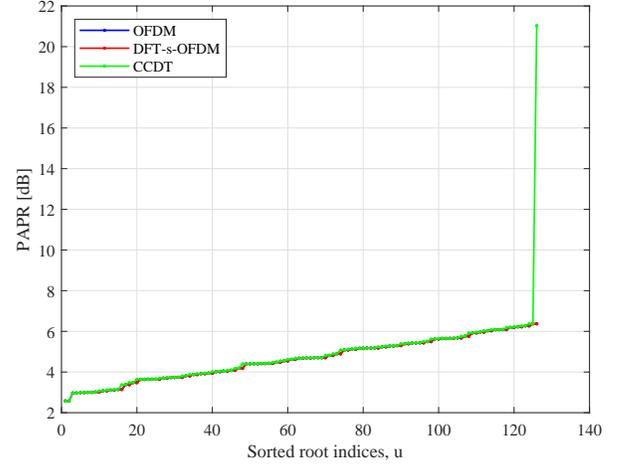}
\caption{PAPRs sorted in increasing order where a ZC sequence of length $N=127$ is used, for OFDM, DFT-s-OFDM and CCDT with $\alpha=2$ and $\beta=1$.}
\label{fig:papr}
\end{center}
\vspace{-0 mm}
\end{figure}

In practice, the PAPR is measured on the time continuous signal. We model this by upsampling $s[n]$, which implies that the PAPR will become larger than the theoretically derived value of 0 dB. Fig. \ref{fig:papr} shows the PAPRs with $Q=4N$, using a ZC sequence of length $N=127$ with different root indices $u$. The PAPRs are displayed in increasing order. With the exception of the PAPR for CCDT with $u=2\alpha$, the PAPRs of the waveforms are comparable. It should be noted that since the DFT of a cyclically-shifted ZC sequence produces the output sequence which is a complex-conjugated and permuted version of the DFT input sequence \cite{Popovic3}. Thus the PAPRs for DFT-s-OFDM are the same as for OFDM.

\section{Performance Evaluations}
\subsection{Range Acquisition}
\begin{table}
\renewcommand{\arraystretch}{1.3}
\caption{List of evaluation parameters and models.}
\vspace{-0.4cm}
\label{table_example}
\begin{center}
\begin{tabular}{c|c}
\hline
\hline
Parameter & Value\\
\hline 
Subcarrier spacing & $f_{\mathrm{SCS}}=15$ kHz\\
\hline
Carrier frequency & $f_c=6$ GHz\\
\hline
Sequence length & $N=127$\\
\hline
Channel model  & \multirow[t]{2}{*} Vehicular A; 0 -- 500 km/h \\
& Clarke's model, $P=5$ paths\\
\hline
\hline
\end{tabular}
\end{center}
\vspace{-0cm}
\end{table}

\begin{table}
\vspace{-0cm}
\renewcommand{\arraystretch}{1.3}
\caption{Multipath channel power delay profile.}
\vspace{-0.4cm}
\label{table_VA}
\begin{center}
\begin{tabular}{c|c|c|c|c|c}
\hline
\hline
$t_l$ [$\mu$s] & 0 & 0.52 & 1.05 & 1.57 & 2.62 \\
\hline
$\mathcal{P}_l$ & 0.4850 &0.4463& 0.0485& 0.0153& 0.0049\\
\hline
\hline
\end{tabular}
\end{center}
\end{table}

We will evaluate timing detection probability of the different waveforms with an m-sequence, i.e., signals having thumbtack-like AF. Additionally, we evaluate CCDT with a ZC sequence and a ridge-like AF according to Property 5, which is robust against Doppler shifts and may improve the detection performance. Timing detection relates to determining range of a target, or acquisition of a synchronization signal. We assume an m-sequence of length $N=127$, which is used as primary synchronization signal sequence in 3GPP NR \cite{38211}. Let us assume a time discrete channel model with the assumptions in Table \ref{table_example} as  
\begin{equation}
h[n]=\sum_{l=0}^{L-1} \sqrt{\mathcal{P}_l}\tilde h_l[n] \delta [n-\tau_l]
\end{equation}
where the relative channel tap powers $\mathcal{P}_l$ and sample delays $\tau_l=t_l/f_s$ of the $L$ taps are obtained from a Vehicular A channel assuming the sampling frequency $f_s=N f_{\mathrm{SCS}}$, according to Table \ref{table_VA}. 
We are considering a time-variant channel using Clarke's two-dimensional isotropic scattering Rayleigh fading model \cite{Clarke}
\begin{equation}
\tilde h_l[n]=\frac{1}{\sqrt{P}} \sum_{p=1}^P e^{j(2\pi f_{\mathrm{D}} n \cos \theta_p+\phi_p)}
\label{eq:clarke}
\end{equation}
where $P$ is the number of propagation paths per channel tap, $f_{\mathrm{D}}=\frac{v}{c}f_c$ is the maximum Doppler frequency, $v$ is the velocity, $c$ is the speed of light, $f_c$ the carrier frequency and $\theta_p$ and $\phi_p$ are the angle of arrival and initial phase of the $p$th propagation path, respectively. Both $\theta_p$ and $\phi_p$ are uniformly distributed over $[-\pi, \pi)$ for all $p$ and they are mutually independent. For $f_{\mathrm{D}}\neq 0$, the channel (\ref{eq:clarke}) varies over a symbol and the subcarriers are no longer orthogonal in the receiver and inter-carrier interference (ICI) occurs.

\begin{figure}
\begin{center}
\includegraphics[width=0.50\textwidth]{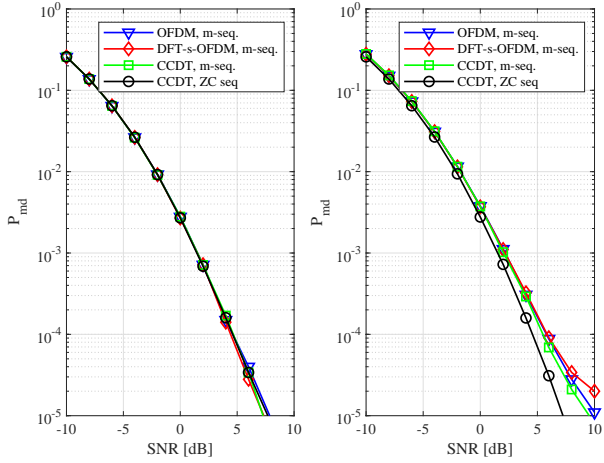}
\caption{Probability of misdetection using an m-sequence, or ZC sequence, of length $N=127$ for 0 km/h (left) and 500 km/h (right) on a Vehicular A channel, for subcarrier spacing $f_{\mathrm{SCS}}=15$ kHz, carrier frequency $f_c=6$ GHz, for the waveforms OFDM, DFT-s-OFDM and CCDT with $\alpha=-2$ and $\beta=-2$.}
\label{fig:missdet}
\end{center}
\vspace{-0 mm}
\end{figure}

\begin{figure}
\begin{center}
\includegraphics[width=0.50\textwidth]{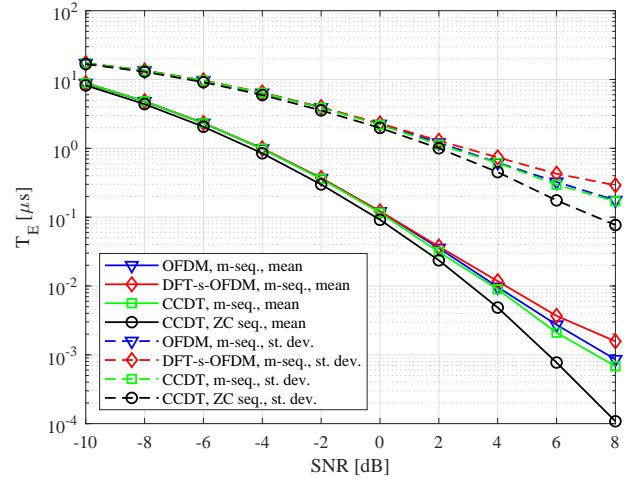}
\caption{Mean and standard deviation of the timing error $T_E$ using an m-sequence, or ZC sequence, of length $N=127$ at 500 km/h on a Vehicular A channel, for subcarrier spacing $f_{\mathrm{SCS}}=15$ kHz, carrier frequency $f_c=6$ GHz, for the waveforms OFDM, DFT-s-OFDM and CCDT with $\alpha=-2$ and $\beta=-2$.}
\label{fig:cdfdet}
\end{center}
\vspace{-0 mm}
\end{figure}

A CP of length $N_{\mathrm{CP}}\ge L$ is attached to $s[n]$ and the received signal $r[n]$, is obtained from convolution with $h[n]$, and adding additive white Gaussian noise (AWGN), $w[n]$. After removing the CP, the signal can be expressed as follows.
\begin{equation}
r[n]=\sum_{l=0}^{L-1} \sqrt{\mathcal{P}_l}\tilde h_l[n] s[n-\tau_l \ (\mathrm{mod}\, N)] +w[n]
\end{equation}
The periodic correlation, which is related to the AF at $\Delta=0$, is then performed as follows to determine the timing sample $\tau^*$.
\begin{align}
\rho(\tau)&=\left | \frac{1}{N}\sum_{n=0}^{N-1} r[n]s^*[n-\tau \ (\mathrm{mod}\, N)] \right | \label{eq:per_corr1}\\
\tau^*&=\arg\max_{\tau \in \{0,1,\ldots,N-1\}} \rho(\tau)
\end{align}
By defining the set of time delay samples of the channel $\mathcal{T}=\{\tau_0,\tau_1,\ldots,\tau_{L-1}\}$, the probability of misdetection, $P_{md}$, is defined by the events of not detecting the received signal on any of the delays in  $\mathcal{T}$.
\begin{equation}
P_{md}=\mathrm{Pr}[\tau^* \not\in  \mathcal{T}] \label{eq:p_md}
\end{equation} 
Determining (\ref{eq:p_md}) on closed-form appears to be a formidable task and we resort to Monte Carlo simulations for its evaluation, for velocities in the range $0 - 500$ km/h. At $f_c=6$ GHz, a velocity of 500 km/h corresponds to a Doppler frequency $f_D=2.78$ kHz or, equivalently, $f_D=0.185 f_{\mathrm{SCS}}$. The effect of the Doppler shift would be the same if $f_{\mathrm{SCS}}$ scales with $f_c$, e.g., using $f_{\mathrm{SCS}}=60$ kHz at $f_c=24$ GHz. The detection (\ref{eq:per_corr1}) is made from one symbol. With an m-sequence, the AFs will be thumbtack-like for all the waveforms, implying that $P_{md}$ should be relatively small and differ moderately between the waveforms.  Fig. \ref{fig:missdet} shows that CCDT performs slightly better than OFDM and DFT-s-OFDM as the velocity increases. However, the detection probability is better for CCDT using a ZC sequence with the parameters chosen to produce an AF with a ridge along the frequency axis, i.e., Fig. \ref{fig:ridges} (b). Such a signal is robust against Doppler shifts and performs better than the other signals when the velocity is large. We define the timing error in the unit of seconds as 
\begin{equation}
T_E=\min_{\tau_l\in\mathcal{T}} \frac{1}{f_s}|\tau^*-\tau_l|
\end{equation}
and estimate the mean and standard deviation of $T_E$ from the simulations. Fig. \ref{fig:cdfdet} confirms the trend of Fig. \ref{fig:missdet}, that the CCDT using a ZC sequence performs slightly better. 

Synchronization signals may need to be detected under large frequency offsets. For example, during initial cell acquisition, prior to when the mobile device has established frequency synchronization with the base station, an oscillator inaccuracy in the order of 10 ppm is typically assumed \cite{38821}. We introduce a frequency offset of $f_\mathrm{o}$ Hz between the transmitter and receiver as 
\begin{equation}
r[n]=\sum_{l=0}^{L-1} \sqrt{\mathcal{P}_l}\tilde h_l[n]e^{j\frac{2\pi}{N}\frac{f_\mathrm{o}}{f_\mathrm{SCS}}n} s[n-\tau_l \ (\mathrm{mod}\, N)] +w[n]
\end{equation}
and signal detection which is based on the AF, i.e., a bank of correlators, each corresponding to a frequency offset hypothesis $\Delta_f$.
\begin{align}
\rho(\tau_f,\Delta_f)&=\left | \frac{1}{N}\sum_{n=0}^{N-1} r[n]s^*[n-\tau_f \ (\mathrm{mod}\, N)]e^{-j\frac{2\pi}{N}\Delta_f n} \right | \label{eq:per_corr}\\
\tau^*&=\arg\max_{\stackrel{\tau_f\in\{0,1,\ldots,N-1\}}{\Delta_f \in \mathcal{D}_H}} \rho(\tau_f,\Delta_f)
\end{align}
We assume that the frequency offset is a uniform random variable $f_\mathrm{o}/f_\mathrm{SCS} \in [-1,1]$ and evaluate $P_{md}$ with $H=3, 5$ or 7 hypotheses, wherein $\mathcal{D}_3=\{-2/3,0,2/3 \}$, $\mathcal{D}_5=\{-4/5,-2/5,0,2/5,4/5 \}$ and $\mathcal{D}_7=\{-6/7,-4/7,-2/7,0,2/7,4/7,6/7 \}$. The number of hypotheses is a trade off between the ability to cancel the frequency offset $f_\mathrm{o}$ and an increase in more false timing candidates. Simulations are made for a range of SNRs and the required SNR to obtain $P_{md}=10^{-4}$ is contained in Table \ref{table_SNR}, which shows that with a velocity of 100 km/h, all schemes perform similarly and there is no gain of using more than 3 hypotheses. With larger velocity, CCDT shows a slight gain and using more hypotheses is better. The last row contains the result for CCDT using a ZC sequence with the parameters chosen to produce an AF with a ridge along the frequency axis, and detection is made without any hyposesis testing, i.e., $\mathcal{D}_1=\{0\}$. Clearly, this signal is insensitive to frequency offsets and outperforms the other signals, with more than 1 dB SNR gain at 100 km/h.

\begin{table*}
\vspace{-0cm}
\renewcommand{\arraystretch}{1.3}
\caption{Required SNR [dB] for $P_{md}=10^{-4}$, for different number of frequency hypotheses using an m-sequence and frequency offset hypothesis testing. Additionally, a ZC sequence is used for CCDT without any frquency offset hypothesis testing.}
\label{table_SNR}
\begin{center}
\begin{tabular}{c|c|c|c|c|c|c|c|c}
\hline
\hline
Velocity & \multicolumn{4}{c|}{100 km/h} & \multicolumn{4}{c}{350 km/h} \\
\hline
Hypotheses & $\mathcal{D}_1$ & $\mathcal{D}_3$ & $\mathcal{D}_5$ &$\mathcal{D}_7$ & $\mathcal{D}_1$ & $\mathcal{D}_3$ & $\mathcal{D}_5$ & $\mathcal{D}_7$\\
\hline
\hline
OFDM & - & 6.20 & 5.95 & 5.95 & - & 6.00 & 5.40 & 5.30\\
\hline
DFT-s-OFDM & - & 6.20 & 5.85 & 5.80 & -& 6.00 & 5.35 & 5.20\\
\hline
CCDT & - & 6.20 & 5.85 & 5.85 & -& 5.70 & 5.25 & 5.05\\
\hline
\hline
CCDT, ZC & 4.6 & - & - & - & 4.7 & - & - & -\\
\hline
\hline
\end{tabular}
\end{center}
\end{table*}

\subsection{Range/Doppler Tracking }
To evaluate the radar properties, we will evalute the different waveforms using thumbtack-like AF produced by either random data symbols, or an m-sequence. We assume a time discrete channel model (cf. \cite{Zeng}) for receiving reflections from $L$ single point targets 
\begin{equation}
h[n]=\sum_{l=0}^{L-1} \sqrt{\mathcal{P}_l}e^{j\phi_l}e^{j\frac{2\pi}{N}\Delta_l n} \delta [n-\tau_l]
\end{equation}
where $\mathcal{P}_l$ are the relative target received powers, $\phi_l$ comprises phase rotations which are uniformly distributed over $[-\pi, \pi)$, $\tau_l$ are the round-trip time delays of the reflected targets and $\Delta_l$ are the Doppler shifts experienced at the receiver due to the motion of the targets. The target speed $v_l$ and range $d_l$ can be determined by $v_l=\frac{f_\mathrm{SCS}\Delta_l c}{2f_c}$ and $d_l=\frac{\tau_l c}{2Nf_\mathrm{SCS}}$. We define the set $\mathcal{P}=\{\mathcal{P}_0,\mathcal{P}_1,\ldots,\mathcal{P}_{L-1}\}$, assume that $\mathcal{P}_0\ge \mathcal{P}_1 \ge \ldots \ge \mathcal{P}_{L-1}$ and that the delays are uniformly distributed from the set $\tau_l \in \{0,1,\ldots,N_{\mathrm{CP}}\}$. The Doppler shift is a continuous random uniform variable with $\Delta_l\in [-1,1]$, i.e., it corresponds to frequencies limited by $\pm f _\mathrm{SCS}$. After removing the CP, the received signal can be described as
\begin{equation}
r[n]=\sum_{l=0}^{L-1} \sqrt{\mathcal{P}_l}e^{j\phi_l}e^{j\frac{2\pi}{N}\Delta_l n} s[n-\tau_l \ (\mathrm{mod}\, N)] +w[n].
\end{equation}
The objective is to estimate the delay $\tau^*$ and Doppler shift $\Delta^*$ for the strongest target, i.e., the other targets are undesired clutter in this respect. Estimation is made by computing a correlation function that is tightly related to the AF. It is evaluated on a 2-D grid of delays and Doppler shifts, which is a common practise \cite{Bhattacharjee}, \cite{Zeng}. Here, the search over Doppler shifts is limited to the set $\mathcal{D}_N=\{-(N-1)/N,-(N-3)/N,\ldots,(N-1)/N \}$.
\begin{align}
\rho(\Delta,\tau)&=\left | \frac{1}{N}\sum_{n=0}^{N-1} r[n]s^*[n-\tau \ (\mathrm{mod}\, N)]e^{-j\frac{2\pi}{N}\Delta n} \right | \label{eq:per_corr2}\\
(\Delta^*,\tau^*)&=\arg\max_{\stackrel{\tau \in \{0,1,\ldots,N_{\mathrm{CP}}\}}{\Delta \in \mathcal{D}_N }} \rho(\Delta,\tau)
\end{align}
\begin{figure}
\begin{center}
\includegraphics[width=0.50\textwidth]{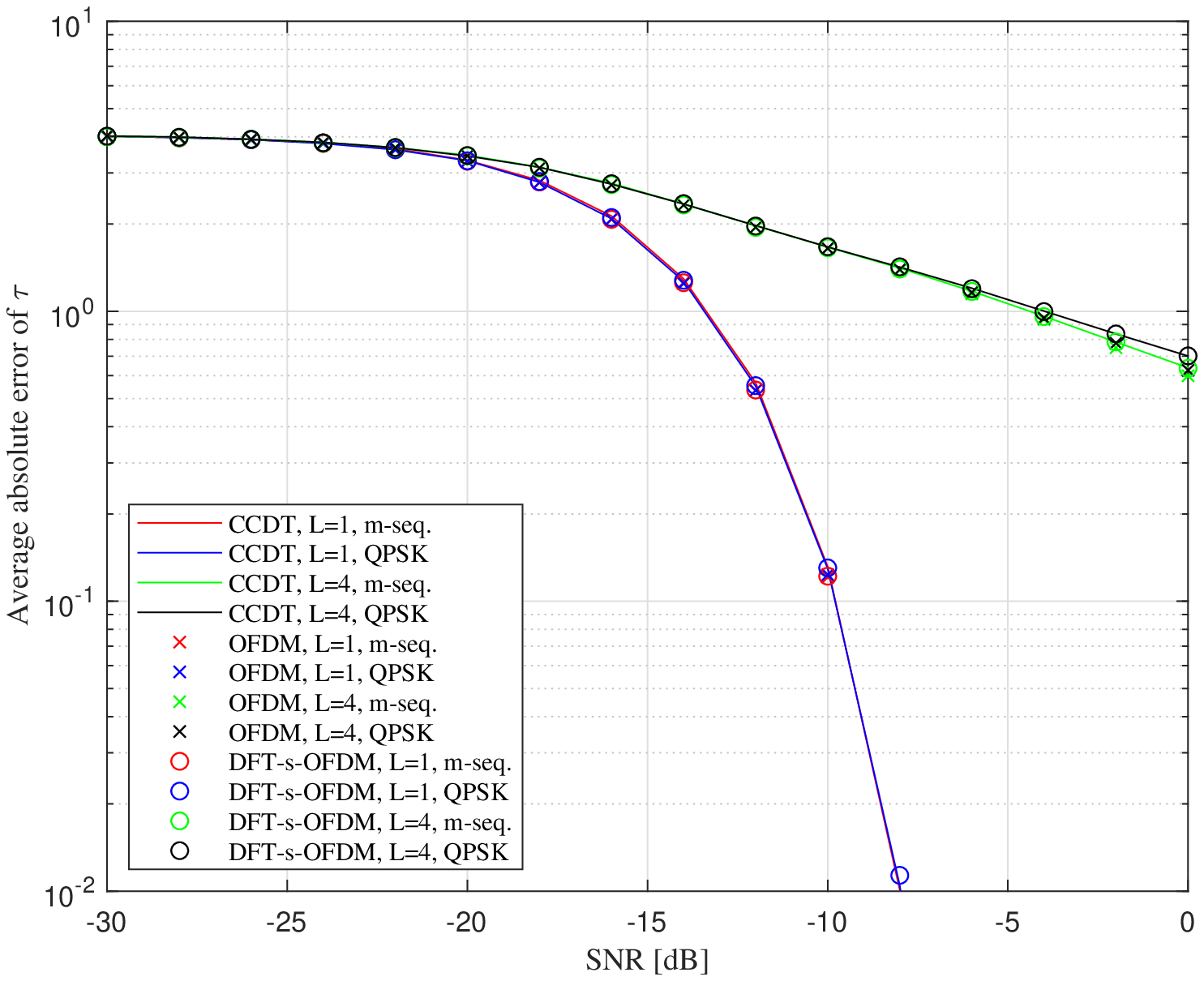}
\caption{Average absolute timing error for different number of targets $L=1$ and $L=4$, false alarm probability $P_{fa}=0.01$, using QPSK or an m-sequence of length $N=127$ for CCDT with $\alpha=-2$ and $\beta=-2$, OFDM and DFT-s-OFDM.}
\label{fig:te}
\end{center}
\vspace{-0 mm}

\begin{center}
\includegraphics[width=0.50\textwidth]{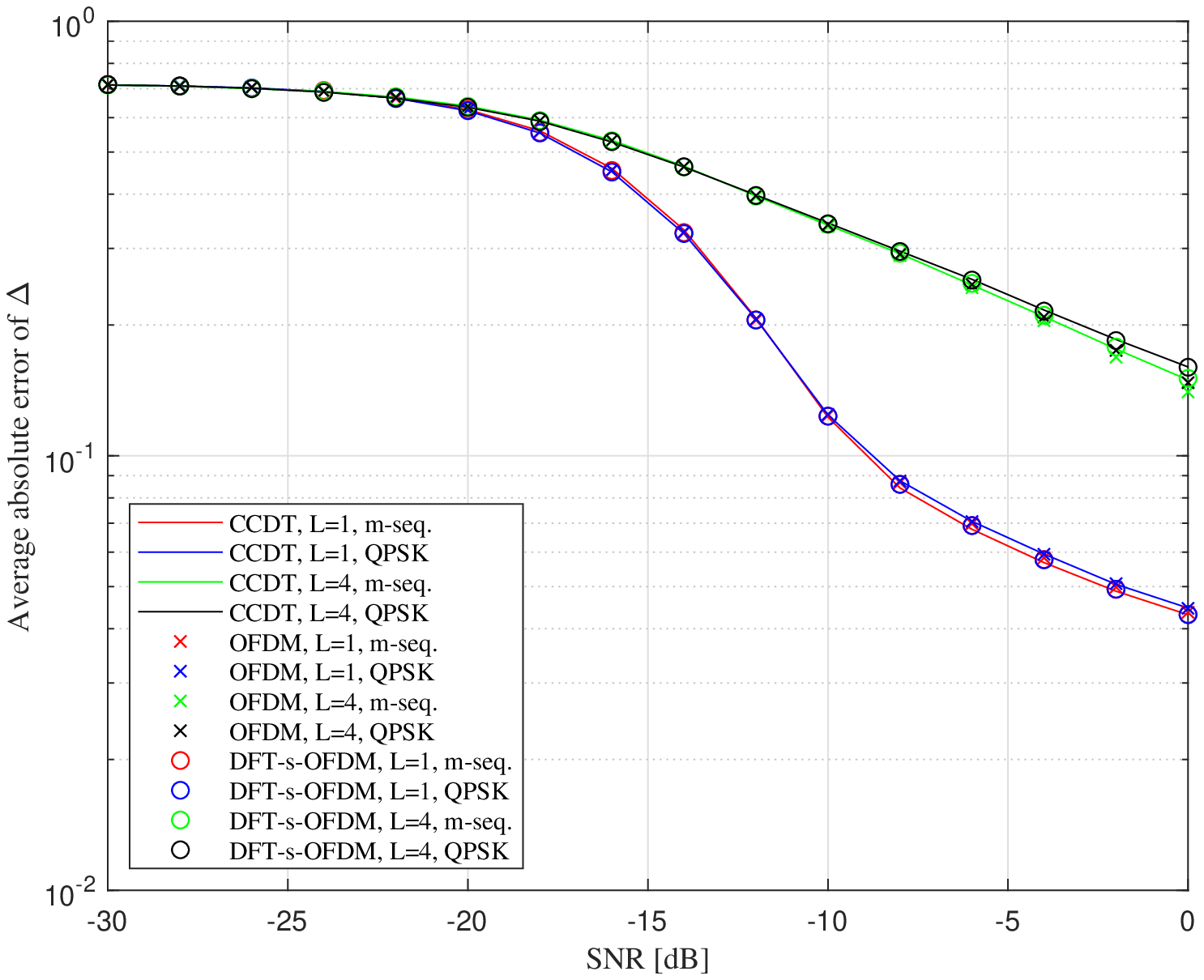}
\caption{Average absolute Doppler shift error for different number of targets $L=1$ and $L=4$,  false alarm probability $P_{fa}=0.01$, using QPSK or an m-sequence of length $N=127$ for CCDT with $\alpha=-2$ and $\beta=-2$, OFDM and DFT-s-OFDM.}
\label{fig:fe}
\end{center}
\vspace{-0 mm}
\end{figure}

The CP length is set to $N_{\mathrm{CP}}=12$ samples. We evaluate two cases: single target ($L=1$) and multiple targets ($L=4$) with $\mathcal{P}=\{1, 0.75, 0.5, 0.25\}$. A detection threshold, $\Gamma$, is determined to control the probability of false alarm, $P_{fa}$, such that $\mathrm{Pr}[\rho(\Delta,\tau)\ge \Gamma]\le P_{fa}$ when the received signal is noise only, $r[n]=w[n]$.  We evaluate by Monte Carlo simulations the average absolute errors for the strongest target, $|\tau^*-\tau_0|$ and $|\Delta^*-\Delta_0|$. Conversion to errors for $d_0$ and $v_0$ can be made as described above for given parameters of $f_\mathrm{SCS}$ and $f_c$. For transmit sequences, we use an m-sequence and a random sequence of QPSK symbols, respectively. As anticipated from Sec. \ref{Sec:random}, both type of sequences exhibit a thumbtack-like AF and the correlation properties are expected to be similar among the sequences and among the waveforms. That is confirmed by Fig. \ref{fig:te} and Fig. \ref{fig:fe}, which shows that m-sequence is only slightly better than a random QPSK sequence for the case with multiple targets. Hence, we can conclude that all these waveforms function as radar signals using the random data QPSK symbols as the transmit sequence. Thereby, spectral efficiency may be improved since dedicated time-frequency resources and a predefined sequence are not needed for a signal serving for radar.  

\section{Conclusions}
It was previously shown that multicarrier chirp waveforms could offer lower bit error rate than OFDM and DFT-s-OFDM \cite{Berggren}\cite{Giroto}\cite{Bhattacharjee}. Herein, it was found that the multicarrier chirp waveform also has gains in detection performance when used as a waveform for synchronization or radar, making it a candidate for JRC. The AF for the multicarrier chirp waveform differs from that of OFDM and DFT-s-OFDM, and was shown to be equal to the AF of the transmit sequence, evaluated at a delay which is a function of $\tau$, $\Delta$ and $\alpha$. The selection of chirp rate, $\alpha$, and transmit sequence can shape the AF to become either thumbtack-like or ridge-like, which are suitable for different applications, e.g., time- and frequency synchronization, or target detection. A signal with ridge-like AF makes it possible to perform time synchronization with no frequency offset hypothesis testing, which otherwise is needed for a signal with a thumbtack-like AF. In particular considering JRC applications, where the modulation symbols are used for both data transmission and radar detection, unitary random modulation symbols produce thumbtack-like AF shape. As in many practical communications systems, including 3GPP cellular standards, high-order quadrature amplitude modulations (QAMs) are used to transmit high-rate date, it remains to be seen whether the thumbtack AFs can be produced with some waveforms modulated by random QAM symbols.


\appendices
\section{Proofs for Property 1 - 3}
\subsection{Lemmas}
\noindent {\it Lemma 1.} For given $N, \Delta$ and $\tau$, and 
$2\alpha k_m = -\Delta + (m-\tau)2\alpha +r_mN$, where $m\in \{0,1,\ldots,N-1\}$ and $k_m\in \{0,1,\ldots,N-1\}$, then
\begin{equation*}
r_m=
\begin{cases}
r_0, & \text{if}\ m<A\\
r_0-2\alpha,&\text{if}\ m\ge A
\end{cases}
\end{equation*} 
where 
\begin{align}
A=&N-k_0 \notag\\
2\alpha k_0 \equiv& -\Delta -2\alpha \tau \ (\mathrm{mod}\, N) \notag\\
r_0=&\frac{2\alpha k_0+\Delta +2\alpha \tau}{N} \notag
\end{align}
\begin{proof}
From (\ref{lincon}), it follows that 
\begin{equation}
k_{m+1}\equiv k_m+1 \ (\mathrm{mod}\, N)
\label{km1}
\end{equation}
since
\begin{equation}
2\alpha k_m = -\Delta + (m-\tau)2\alpha +r_mN
\label{eq:womod}
\end{equation}
implies that 
\begin{equation}
2\alpha k_m = -\Delta + (m-\tau)2\alpha \ (\mathrm{mod}\, N)
\label{eq:wmod}
\end{equation}
and
\begin{align}
2\alpha k_{m+1}=& -\Delta + 2\alpha (m+1) - 2 \alpha \tau \ (\mathrm{mod}\, N) \notag \\
=& 2\alpha k_m +2\alpha \ (\mathrm{mod}\, N).
\end{align}
From (\ref{km1}), the elements of the sequence $k_m$ belong to the set $k_m \in \{0,1,\ldots, N-1\}$ and can be expressed as
\begin{equation}
k_{m+1}=
\begin{cases}
k_m+1, & \text{if}\ k_m+1<N\\
k_m+1-N, & \text{if}\ k_m+1 = N. 
\end{cases}
\label{km+1a}
\end{equation}
The case $k_m+1 = N$ will occur for a single value of $m=A-1$, because then $k_{m+1} \equiv 0  \ (\mathrm{mod}\, N)$, and 0 appears only once in the sequence $k_m$. Let us assume that $A$ is a positive integer such that $A=N-k_0$, where $k_0$ satisfies (\ref{eq:wmod}) as
\begin{equation}
2\alpha k_0 = -\Delta -2\alpha \tau \ (\mathrm{mod}\, N).
\label{k0}
\end{equation}
Then from (\ref{eq:wmod}), (\ref{k0}) and using $2\alpha N \ (\mathrm{mod}\, N)=0$ since $2\alpha\in \mathbb{Z}$ \cite{Berggren}, we have 
\begin{align}
2\alpha k_A \equiv & -\Delta +(A-\tau)2\alpha \ (\mathrm{mod}\, N) \notag\\
\equiv&  -\Delta -\tau2\alpha-k_02\alpha \ (\mathrm{mod}\, N)\notag \\
\equiv&  -\Delta -\tau2\alpha-(-\Delta -\tau2\alpha) \ (\mathrm{mod}\, N)\notag \\
\equiv& 0 \notag
\end{align}
and thus rewrite (\ref{km+1a}) as 
\begin{equation}
k_{m+1}=
\begin{cases}
k_m+1, & \text{if}\ m\neq A-1\\
k_m+1-N, & \text{if}\ m=A-1. 
\end{cases}
\label{km+1}
\end{equation}
The corresponding values can be obtained by inserting the values of $k_{m+1}$ from (\ref{km+1}) into (\ref{eq:wmod}). Thus, for $m\neq A-1$ we obtain $r_{m+1}=r_m$ and for $m=A-1$ it follows that $r_{m+1}=r_m-2\alpha$. Since $2\alpha \in \mathbb{Z}$, it follows that $r_{m+1}\in \mathbb{Z}$.  Hence, we have
\begin{align}
r_0&=r_1=\ldots=r_{A-1}\notag\\  
r_{N-1}&=r_{N-2}=\ldots=r_A=r_{A-1}-2\alpha \notag
\end{align}
and obtain 
\begin{equation}
r_m=
\begin{cases}
r_0, & \text{if}\ m< A-1\\
r_0-2\alpha, & \text{if}\ m\ge A-1. 
\end{cases}
\label{km+1b}
\end{equation}
Using (\ref{eq:womod}), it follows that 
\begin{equation*}
r_0=\frac{2\alpha k_0+\Delta +2\alpha \tau}{N}.
\end{equation*}
\end{proof}

\noindent {\it Lemma 2.} For any $0\le m \le N-1$ and integer $\Delta$, $e^{j\pi\frac{\Delta r_m}{\alpha}}=e^{j\pi\frac{\Delta r}{\alpha}}$, for any $r$ in the set $r\in \{r_0,r_1,\ldots,r_{N-1}\}$.
\begin{proof}
From Lemma 1, it follows that there are at most two different values of $r_m$, i.e., $r_m=r_0$ or $r_m=r_0-2\alpha$. Since $e^{j\pi\frac{\Delta (r_0-2\alpha)}{\alpha}}=e^{j\pi\frac{\Delta r_0}{\alpha}}$, the Lemma follows.
\end{proof}

\noindent {\it Lemma 3.} For any $0\le m \le N-1$, $e^{-j\frac{2\pi}{N}\left( \frac{r^2_mN^2}{4\alpha}-\beta \frac{r_mN}{2\alpha} \right )}=e^{-j\frac{2\pi}{N}\left( \frac{r^2_0N^2}{4\alpha}-\beta \frac{r_0N}{2\alpha} \right )}$.
\begin{proof}
From Lemma 1, it follows that there are at most two different values of $r_m$, i.e., $r_m=r_0$ or $r_m=r_0-2\alpha$. Thus, if $r_m=r_0-2\alpha$ we have 
\begin{align}
e^{-j\frac{2\pi}{N}\left( \frac{r^2_mN^2}{4\alpha}-\beta \frac{r_mN}{2\alpha} \right )}=&e^{-j\frac{2\pi}{N}\left( \frac{(r_0-2\alpha)^2N^2}{4\alpha}-\beta \frac{(r_0-2\alpha)N}{2\alpha} \right )} \notag \\
=& e^{-j\frac{2\pi}{N}\left( \frac{r^2_0N^2}{4\alpha}-\beta \frac{r_0N}{2\alpha} \right )}
e^{j2\pi r_0N} e^{-j2\pi(\alpha N+\beta)} \notag \\
=& e^{-j\frac{2\pi}{N}\left( \frac{r^2_0N^2}{4\alpha}-\beta \frac{r_0N}{2\alpha} \right )}
\end{align}
where the last step follows from (\ref{periodicity}).
\end{proof}

\subsection{Proof for Property 1}
Inserting (\ref{ccdt}) in (\ref{peramb}) and utilizing that (\ref{periodicity}) makes (\ref{basis}) to have a period $N$ \cite{Berggren}, i.e., $g[k]=g[k\ (\mathrm{mod} \,N)]$, we obtain 
\begin{align}
\chi (\Delta,\tau)&= \frac{1}{N} \sum_{n=0}^{N-1}\sum_{m=0}^{N-1}\sum_{k=0}^{N-1}x[k]g[n-k] \notag \\
 &\times x^*[m]g^*[n+\tau-m]  e^{j\frac{2\pi}{N}\Delta n}\notag\\
&= \sum_{m=0}^{N-1}\sum_{k=0}^{N-1} x[k]x^*[m] e^{-j\frac{2\pi}{N}\left(\alpha\left(k^2-m^2\right ) +\beta (m-k)+2\alpha\tau m\right )} \notag \\
 &\times C_0 \sum_{n=0}^{N-1}  e^{j\frac{2\pi}{N}n \left (\Delta - 2\alpha (m-k)+2\alpha \tau \right)}. 
\label{amb1}
\end{align}
with $C_0=\frac{g^*[\tau]e^{-j\frac{2\pi}{N}\gamma}}{N\sqrt{N}}$.
A well-known identity is that for $k\in \mathbb{Z}$: 
\begin{equation}
\sum_{n=0}^{N-1} e^{j\frac{2\pi}{N}nk}=N\delta[k\ (\mathrm{mod} \,N)]
\label{geosum}
\end{equation}
Therefore, the inner sum in (\ref{amb1}) is non-zero and equal to $N$ only when there for each $m$ exists a $k=k_m$ and $k \in \{0,1,\ldots, N-1\}$ such that 
\begin{equation}
2\alpha k_m \equiv -\Delta + (m-\tau)2\alpha \ (\mathrm{mod} \,N).
\label{lincon}
\end{equation}
The condition (\ref{lincon}) is a linear congruence equation having only a single $k_m$ as solution for each $m$ due to (\ref{orthogonality}). We can rewrite (\ref{lincon}) as
\begin{equation}
2\alpha k_m = -\Delta + (m-\tau)2\alpha +r_mN
\label{criterion}
\end{equation}
where $r_m\in \mathbb{Z}$. Inserting $k_m$ from (\ref{criterion}) in (\ref{amb1}) gives
\begin{align}
\chi (\Delta,\tau)&=C_1 \sum_{m=0}^{N-1} x\left [m-\tau+\frac{r_mN-\Delta}{2\alpha}\right]x^*[m]  \notag\\
& \times e^{-j\frac{2\pi}{N}\left( \frac{r^2_mN^2}{4\alpha}-\beta \frac{r_mN}{2\alpha} \right )} e^{j\pi\frac{\Delta r_m}{\alpha}} 
e^{j\frac{2\pi}{N}\Delta m}  \notag\\
&\stackrel{(\mathrm{a})}{=}C  \sum_{m=0}^{N-1} x\left [m-\tau+\frac{r_mN-\Delta}{2\alpha}\right]x^*[m]e^{j\frac{2\pi}{N}\Delta m} \notag \\
&\stackrel{(\mathrm{b})}{=}C  \sum_{m=0}^{N-1} x\left [m-\tau+\frac{r_0N-\Delta}{2\alpha} \ (\mathrm{mod} \,N)\right]x^*[m] \notag \\
& \times e^{j\frac{2\pi}{N}\Delta m} 
\label{amb3}
\end{align}
with $C_1=C_0\sqrt{N}e^{j\frac{2\pi}{N}\gamma}g^*[\tau+\Delta/2\alpha] $ and $C=C_1 N\sqrt{N}e^{j\frac{2\pi}{N}\gamma}g^*[-r_0N/2\alpha]e^{j\pi\frac{\Delta r_0}{\alpha}}$, where replacing $r_m$ with $r_m=r_0$ and moving the exponential terms outside the sum in (a) follows from Lemma 2 and Lemma 3. Replacing $r_m$ with $r_m=r_0$ or $r_m=r_0-2\alpha$ in the sequence argument in (b) follows from Lemma 1, and since $2\alpha$ is an integer \cite{Berggren} and $-2\alpha N\equiv 0 \ (\mathrm{mod} \,N)$, the modulo-$N$ operator is introduced.

\subsection{Proof for Property 2}
The condition $\Delta +2\alpha \tau \equiv 0 \ (\mathrm{mod} \,N)$ is equivalent to $\Delta +2\alpha \tau=rN$ with $r\in\mathbb{Z}$, which is obtained from (\ref{criterion}) by using $k_m=m$ and $r_m=r$. Therefore, $(r_mN-\Delta)/2\alpha=\tau$ and from (\ref{eq:ambfnl}), utilizing $|C|=1/N$ and (\ref{geosum}) we have 
\begin{equation}
|\chi (\Delta,\tau)|=\frac{1}{N}\left | \sum_{m=0}^{N-1} x[m]x^*[m] e^{j\frac{2\pi}{N}\Delta m} \right |=
\begin{cases}
0, & \Delta\neq 0\\
1, & \Delta= 0.
\end{cases}
\end{equation} 
It follows from (\ref{orthogonality}) that if $\Delta +2\alpha \tau \equiv 0 \ (\mathrm{mod} \,N)$ and $\Delta=0$, then $\tau=0$, i.e., the AF is $|\chi (\Delta,\tau)|=\delta[\Delta]$.
 
\subsection{Proof of Property 3}
Let us define $\epsilon(\tau)=-\tau+(r_0N-\Delta)/2\alpha$ and insert $x[m]$ in (\ref{eq:ambfnl}) to obtain
\begin{equation}
\begin{aligned}
\sum_{\tau=0}^{N-1}|\chi (\Delta,\tau)|^2=&\sum_{\tau=0}^{N-1}\chi (\Delta,\tau)\chi^*(\Delta,\tau) \notag\\
\stackrel{(\mathrm{a})}{=}&\frac{1}{N^2}\sum_{\tau=0}^{N-1} \sum_{n=0}^{N-1}x[n]x^*[n+\epsilon(\tau)]e^{j\frac{2\pi}{N}\Delta n} \notag\\
\times & \sum_{m=0}^{N-1}x^*[m]x[m+\epsilon(\tau)]e^{-j\frac{2\pi}{N}\Delta m} \notag\\
\stackrel{(\mathrm{b})}{=}&\frac{1}{N^2}\sum_{n=0}^{N-1} \sum_{m=0}^{N-1}x[n]x^*[m]e^{j\frac{2\pi}{N}\Delta (n-m)} \notag \\
\times & \sum_{\tau=0}^{N-1}x^*[n+\epsilon(\tau)]x[m+\epsilon(\tau)] \notag \\
\stackrel{(\mathrm{c})}{=}& \frac{1}{N^2}\sum_{n=0}^{N-1} \sum_{m=0}^{N-1}x[n]x^*[m]e^{j\frac{2\pi}{N}\Delta (n-m)} \notag \\
\times & N\delta[n-m]\\
\stackrel{(\mathrm{d})}{=}&\frac{1}{N}\sum_{n=0}^{N-1} x[n]x^*[n] \notag \\
\stackrel{(\mathrm{e})}{=}& 1
\end{aligned}
\end{equation}
where (a)-(b) follow by definition, (c) is due to (\ref{zac}) and (d)-(e) are due to (\ref{ca}), and all additions of sequence indices in $x[m]$ are performed $(\mathrm{mod} \,N)$.

\section{Proofs for Property 4 - 6}
\subsection{Proof for Property 4}
By insertion of $x_k[m]$ in (\ref{eq:ambfnl}) and (\ref{geosum})
\begin{align}
|\chi (\Delta,\tau)|&=\frac{1}{N} \left | \sum_{m=0}^{N-1} e^{j\frac{2\pi}{N} \frac{kr_mN}{2\alpha}} e^{j\frac{2\pi}{N}\Delta m} \right | \notag\\
&=\frac{1}{N} \left | \sum_{m=0}^{N-1}e^{j\frac{2\pi}{N}\Delta m} \right | =
\begin{cases}
0, & \Delta\neq 0\\
1,& \Delta=0. 
\end{cases}
\end{align}
where it follows from Lemma 1 in Appendix A, that $e^{j\frac{2\pi}{N} \frac{kr_mN}{2\alpha}}=e^{j\frac{2\pi}{N} \frac{kr_0N}{2\alpha}}$, because there are at most two different values of $r_m$, i.e., $r_0$ or $r_0-2\alpha$, and $e^{j\frac{2\pi}{N} \frac{k(r_0-2\alpha)N}{2\alpha}}=e^{j\frac{2\pi}{N} \frac{kr_0N}{2\alpha}}$.

\subsection{Proof for Property 5}
Let us define $\epsilon=-\tau+(r_0N-\Delta)/2\alpha$ and insert $x_u[m]$ in (\ref{eq:ambfnl}) to obtain
\begin{equation}
|\chi (\Delta,\tau)|=\frac{1}{N} \left | \sum_{m=0}^{N-1} e^{j\frac{2\pi}{N}m(\Delta+u\epsilon)} \right |
\label{zcamb}
\end{equation}
where we have used $x_u[m+\epsilon \ (\mathrm{mod} \,N) ]=x_u[m+\epsilon]$. The modulus AF (\ref{zcamb}) is equal to $1$ when 
\begin{equation*}
\Delta+u\left (-\tau+\frac{r_0N-\Delta}{2\alpha} \right) \equiv 0 \ (\mathrm{mod} \,N)
\end{equation*}
which can be simplified as 
\begin{equation}
2\alpha \Delta-u\left (2\alpha\tau+\Delta \right) \equiv 0 \ (\mathrm{mod} \,N).
\label{zccond}
\end{equation}
When $u=2\alpha$ and $\tau=0$, (\ref{zccond}) holds for all $\Delta$, i.e., $|\chi (\Delta,\tau=0)|=1$. When $u=2\alpha$, $\tau\neq 0$ and $\Delta \neq 0$, we have   
$2\alpha \Delta-2\alpha\left (2\alpha\tau+\Delta \right)=-4\alpha^2\tau$ and due to (\ref{orthogonality}) and that $\tau$ is an integer, it follows that $-4\alpha^2\tau\nequiv 0 \ (\mathrm{mod} \,N)$, and thus, $|\chi (\Delta \neq 0,\tau \neq 0)|=0$.

A similar property can be found when $N$ is even and $x_u[m]=e^{j\frac{\pi}{N} u m^2}$, which is omitted here for brevity. Since a ZC sequence  is a CAZAC sequence, Property 5 could alternatively be proven using Property 3 and noting that when $u=2\alpha$ and $\tau=0$, (\ref{zccond}) holds for all $\Delta$, i.e., $|\chi (\Delta,\tau=0)|=1$, and thus  $|\chi (\Delta,\tau \neq 0)|=0$. 
\subsection{Proof for Property 6}
\noindent{\it Lemma 4} If $x[m]$ is an m-sequence and $0<\tau\le N-1$, then $x[m]x[m+\tau\ (\mathrm{mod}\, N)]=-x[m+\tau'\ (\mathrm{mod}\, N)]$ for some $\tau'$.
\begin{proof}
Let $y[m]$ be a binary m-sequence and $x[m]=q[y[m]]$ with 
\begin{equation}
q[b]=
\begin{cases}
-1, & b=0\\
1, & b=1.
\end{cases}
\end{equation}
It is straightforward to verify that for $b_0\in\{0,1\}$ and $b_1\in\{0,1\}$:
\begin{equation}
q(b_0+b_1 \ (\mathrm{mod}\, 2))=-q(b_0)q(b_1)
\label{qmap}
\end{equation}
The shift-and-add property of m-sequences gives that $y[m+\tau\ (\mathrm{mod}\, N)+y[m]\ (\mathrm{mod}\, 2)=y[m+\tau'\ (\mathrm{mod}\, N)]$ for $0<\tau\le N-1$, $m=0,1,\ldots,N-1$, where $\tau'$ depends on $\tau$. Therefore, applying (\ref{qmap}) to this identity results in $x[m]x[m+\tau\ (\mathrm{mod}\, N)]=-x[m+\tau'\ (\mathrm{mod}\, N)]$.
\end{proof}

\noindent {\it Lemma 5.} If $x[m]$ is an m-sequence and  $|\chi (\Delta\neq 0,\tau_0)|\neq 0$ and if $\Delta+2\alpha\tau_1\nequiv 0 \ (\mathrm{mod} \,N)$, then $|\chi (\Delta\neq 0,\tau_1)|=|\chi (\Delta\neq 0,\tau_0)|$ for $\tau_0\neq \tau_1$.
\begin{proof}
Let us define $\epsilon(\tau)=-\tau+(r_0N-\Delta)/2\alpha$. From (\ref{criterion}), by using $k_m=m$ there exists an $r_m$ such that $\Delta=-2\alpha\tau+rN$, thus $\epsilon(\tau)\nequiv 0 \ (\mathrm{mod} \,N)$. Consequently, if  $\Delta+2\alpha\tau\nequiv 0 \ (\mathrm{mod} \,N)$, then $\epsilon(\tau)\nequiv 0 \ (\mathrm{mod} \,N)$. Assume that $|\chi (\Delta\neq 0,\tau_0)|\neq 0$ and an integer $t$ such that
\begin{align}
|\chi (\Delta\neq 0,\tau_0)|&=\frac{1}{N}\left |\sum_{m=0}^{N-1} \!\! x[m+\epsilon(\tau_0) \ (\mathrm{mod} \,N)]x[m]e^{j\frac{2\pi}{N}\Delta m} \right | \notag \\
&\stackrel{\mathrm{(a)}}{=}\frac{1}{N}\left |\sum_{m=0}^{N-1} x[m+\tau' \ (\mathrm{mod} \,N)]e^{j\frac{2\pi}{N}\Delta m} \right | \notag \\
&\stackrel{\mathrm{(b)}}{=} \frac{1}{N}\Bigg |\sum_{m=t}^{N-1+t}  x[m+\tau' -t\ (\mathrm{mod} \,N)] \notag \\
&\times e^{j\frac{2\pi}{N}\Delta (m-t)} \Bigg |\notag \\
&\stackrel{\mathrm{(c)}}{=} \frac{1}{N}\left |\sum_{m=0}^{N-1} \!\! x[m+\tau' -t\ (\mathrm{mod} \,N)]e^{j\frac{2\pi}{N}\Delta (m-t)}\right | \notag \\
&\stackrel{\mathrm{(d)}}{=} \frac{1}{N}\Bigg |\sum_{m=0}^{N-1}  x[m+\tau_1\ (\mathrm{mod} \,N)]x^*[m] \notag \\
& \times e^{j\frac{2\pi}{N}\Delta (m-t)}\Bigg | \notag \\
&\stackrel{\mathrm{(e)}}{=}\left |\chi (\Delta\neq 0,\tau_1)e^{-j\frac{2\pi}{N}\Delta t}\right | \notag \\
&\stackrel{\mathrm{(f)}}{=}\left |\chi (\Delta\neq 0,\tau_1) \right |   
\end{align}
where Lemma 4 was used in (a) and (d) and the change of summation index from (b) to (c) follows from the periodicity in $N$ of the exponential function and the sequence $x[m]$. 
\end{proof}

\noindent {\it Lemma 6.} If $x[m]$ is an m-sequence, then
\begin{align}
\sum_{\tau=0}^{N-1}|\chi (\Delta,\tau)|^2=
\begin{cases}
1+(N-1)\frac{1}{N^2}, & \Delta=0\notag \\
1-\frac{1}{N^2}, & \Delta\neq 0.\notag \\
\end{cases}
\end{align}
\begin{proof}
Let us define $\epsilon(\tau)=-\tau+(r_0N-\Delta)/2\alpha$ and insert $x[m]$ in (\ref{eq:ambfnl}) to obtain
\begin{equation}
\begin{aligned}
\sum_{\tau=0}^{N-1}|\chi (\Delta,\tau)|^2=&\sum_{\tau=0}^{N-1}\chi (\Delta,\tau)\chi^*(\Delta,\tau) \notag\\
=&\frac{1}{N^2}\sum_{\tau=0}^{N-1} \sum_{n=0}^{N-1}x[n]x[n+\epsilon(\tau)]e^{j\frac{2\pi}{N}\Delta n} \notag\\
\times & \sum_{m=0}^{N-1}x[m]x[m+\epsilon(\tau)]e^{-j\frac{2\pi}{N}\Delta m} \notag\\
=&\frac{1}{N^2}\sum_{n=0}^{N-1} \sum_{m=0}^{N-1}x[n]x[m]e^{j\frac{2\pi}{N}\Delta (n-m)} \notag \\
\times & \sum_{\tau=0}^{N-1}x[n+\epsilon(\tau)]x[m+\epsilon(\tau)] \notag \\
=&\frac{1}{N^2}\sum_{n=0}^{N-1} \sum_{m=0}^{N-1}x[n]x[m]e^{j\frac{2\pi}{N}\Delta (n-m)} \notag \\
\times & N\rho(n-m)  \notag \\
=& \frac{1}{N}\sum_{t=0}^{N-1} \sum_{m=0}^{N-1}x[m+t]x[m]e^{j\frac{2\pi}{N}\Delta t}\rho(t)  \notag \\
=& \sum_{t=0}^{N-1} \rho^2(t)e^{j\frac{2\pi}{N}\Delta t}   \\
\label{eqh}
\end{aligned}
\end{equation}
where all additions of sequence indices are performed $(\mathrm{mod} \,N)$ and where the change to summation index $t$ follows from the periodicity in $N$ of the exponential function and the periodicity of the sequence $x[m \ (\mathrm{mod} \,N)]$.
If $\Delta=0$, then $\sum_{t=0}^{N-1} \rho^2(t)e^{j\frac{2\pi}{N}\Delta t}=1+(N-1)\frac{1}{N^2}$. If $\Delta\neq 0 \ (\mathrm{mod} \,N)$, then 
\begin{align}
\sum_{t=0}^{N-1} \rho^2(t)e^{j\frac{2\pi}{N}\Delta t}=&\rho^2(0)+\sum_{t=1}^{N-1} \rho^2(t)e^{j\frac{2\pi}{N}\Delta t} \notag \\
=&1+\frac{1}{N^2}\left (\sum_{t=0}^{N-1} e^{j\frac{2\pi}{N}\Delta t}-\rho^2(0) \right) \notag \\
=&1+\frac{1}{N^2}(0-1) \notag \\
=&1-\frac{1}{N^2}. 
\end{align}
Thus, 
\begin{equation}
\sum_{\tau=0}^{N-1}|\chi (\Delta,\tau)|^2=
\begin{cases}
1+(N-1)\frac{1}{N^2}, & \Delta=0\notag \\
1-\frac{1}{N^2}, & \Delta\neq 0. \\
\end{cases}
\end{equation}
\end{proof}

Property 6 can then be proven as follows. The first case is trivial and case two and three follow straightforwardly from (\ref{corrmseq}) and Property 2, respectively. The fourth case is proven as follows. For a given $\Delta\neq 0$, Property 2 gives that there exists one $\tau \ (\tau \in \{0,1,\ldots,N-1\})$ for which $|\chi (\Delta\neq 0,\tau)|= 0$, since the linear congruence equation $\Delta+2\alpha\tau  \equiv 0 (\mathrm{mod} \,N)$ has one solution $\tau$ when (\ref{orthogonality}) holds. Thus, are $N-1$ values of $\tau$ where $|\chi (\Delta\neq 0,\tau)|\neq 0$. From Lemma 5 and Lemma 6, we then have $(N-1)|\chi (\Delta\neq 0,\tau)|^2=1-1/N^2$ and we can solve for $|\chi (\Delta\neq 0,\tau)|=\sqrt{(N+1)}/N$.

\section{Ambiguity Function with Upsampling}
Consider (\ref{eq:ccdt_fd}) with upsampling such that $(Q/N)\in \mathbb{Z}$ and
\begin{equation}
s[n]=\frac{1}{\sqrt{Q}} \sum_{m=0}^{N-1} G[m]X[m]e^{j\frac{2\pi}{Q}mn}
\label{eq:upsamp}
\end{equation}
for $n=0,1,\ldots,Q-1$. Using (\ref{eq:upsamp}), (\ref{eq:G}) and (\ref{eq:X}), we obtain
\begin{align}
\chi (\Delta,\tau)=&\frac{1}{Q}\sum_{n=0}^{Q-1} s[n]s^*[n+\tau \ (\mathrm{mod}\, Q)]e^{j\frac{2\pi}{Q}\Delta n} \notag \\
=& \frac{1}{Q^2} \sum_{m=0}^{N-1}\sum_{p=0}^{N-1} G[m]X[m]G^*[p]X^*[p]e^{-j\frac{2\pi}{Q}p\tau} \notag \\
 \times & \sum_{n=0}^{Q-1}e^{j\frac{2\pi}{Q}n(m-p+\Delta)} \label{eq:upgen}
\end{align}
which holds for arbitrary $\Delta$. For non-integer $\Delta$, the inner sum in (\ref{eq:upgen}) can be replaced by:
\begin{align}
\sum_{n=0}^{Q-1}e^{j\frac{2\pi}{Q}n(m-p+\Delta)}=&\frac{\sin \left ( \pi (m-p+\Delta) \right )}{\sin \left (\frac{\pi (m-p+\Delta)}{Q} \right )} \notag\\
\times & e^{j\frac{\pi (Q-1)(m-p+\Delta)}{Q}} \label{eq:upgen2}
\end{align}
For integer $\Delta$, the inner sum can be replaced by $Q\delta[m-p+\Delta \ (\mathrm{mod}\, Q)]$ and we can proceed from (\ref{eq:upgen}) by 
\begin{align}
\chi (\Delta,\tau)=& \frac{1}{Q} \sum_{p=0}^{N-1} G[p-\Delta+r_pQ]X[p-\Delta+r_pQ] \notag \\
 \times & G^*[p]X^*[p]e^{-j\frac{2\pi}{Q}p\tau} \label{eq:chiu}
\end{align}
 where $r_p\in \mathbb{Z}$. Furthermore, using (\ref{eq:G}) and (\ref{eq:X}), it follows that 
 \begin{align}
 G[p-\Delta+r_pQ]G^*[p]=&\sum_{k=0}^{N-1}  g[k]e^{-j\frac{2\pi}{N}(p-\Delta+r_pQ)k} \notag \\
 \times & \sum_{t=0}^{N-1}  g^*[t]e^{j\frac{2\pi}{N}pt} \notag \\
 =& \sum_{v=0}^{N-1} \sum_{k=0}^{N-1}g[k]g^*[k+v] \notag \\
 \times & e^{j\frac{2\pi}{N}pv}e^{j\frac{2\pi}{N}\Delta k} \notag \\
 =& N \sum_{v=0}^{N-1} \chi_g (\Delta,v)e^{j\frac{2\pi}{N}pv}
\end{align}
and similarly
\begin{align}
 X[p-\Delta+r_pQ]X^*[p]=&\frac{1}{N}\sum_{k=0}^{N-1}  x[k]e^{-j\frac{2\pi}{N}(p-\Delta+r_pQ)k} \notag \\
  \times & \sum_{t=0}^{N-1}  x^*[t]e^{j\frac{2\pi}{N}pt} \notag \\
  =&\sum_{w=0}^{N-1} \chi_x (\Delta,w)e^{j\frac{2\pi}{N}pw}.
\end{align}
Therefore, by using (\ref{eq:geosum2}), (\ref{eq:chiu}) can be written as: 
\begin{align}
\chi (\Delta,\tau)=&\frac{N}{Q} \sum_{v=0}^{N-1} \chi_g (\Delta,v) \sum_{w=0}^{N-1} \chi_x (\Delta,w)\notag \\
\times & \sum_{p=0}^{N-1} e^{j\frac{2\pi}{N}p(v+w-\frac{N}{Q}\tau)} \label{eq:chiup}\\
= &\frac{N}{Q} \sum_{v=0}^{N-1} \sum_{w=0}^{N-1}\chi_g (\Delta,v)  \chi_x (\Delta,w)\notag \\
 \times &\frac{\sin \left (\pi \left (v+w-\frac{N}{Q}\tau \right ) \right ) }{\sin \left ( \frac{\pi \left (v+w-\frac{N}{Q}\tau \right ) }{N}\right )}
e^{j\frac{\pi(N-1)\left (v+w-\frac{N}{Q}\tau \right )}{N}}.\label{eq:ambups}
\end{align}
For the special case of $Q=N$, i.e., no upsampling, (\ref{geosum}) can be used to give $N\delta[v+w-\tau \ (\mathrm{mod}\, N)]$ for the inner sum in (\ref{eq:chiup}) such that 
\begin{align}
\chi (\Delta,\tau)=&N \sum_{v=0}^{N-1} \chi_g (\Delta,v) \chi_x (\Delta,\tau-v). \label{eq:ambconv}
\end{align}

\section{Alternative derivation of Property 1}
By using (\ref{basis}), it follows that
\begin{align}
\chi_g (\Delta,\tau)=&\frac{1}{N}\sum_{n=0}^{N-1} g[n]g^*[n+\tau \ (\mathrm{mod}\, N)]e^{j\frac{2\pi}{N}\Delta n} \notag \\
=& \frac{1}{N^2}e^{j\frac{2\pi}{N}(\alpha \tau^2 +\beta \tau)}\sum_{n=0}^{N-1}e^{j\frac{2\pi}{N}(2\alpha \tau+\Delta)n} \notag\\
=&\frac{1}{N}e^{j\frac{2\pi}{N}(\alpha \tau^2 +\beta \tau)}\delta[2\alpha \tau+\Delta \ (\mathrm{mod}\, N)]. \label{eq:chig}
\end{align}
Furthermore, by definition it follows that 
\begin{align}
\chi_x (\Delta,\tau)=&\frac{1}{N}\sum_{n=0}^{N-1} x[n]x^*[n+\tau \ (\mathrm{mod}\, N)]e^{j\frac{2\pi}{N}\Delta n} \notag \\
=& \frac{1}{N}\sum_{n=0}^{N-1} x[n-\tau \ (\mathrm{mod}\, N)]x^*[n]e^{j\frac{2\pi}{N}\Delta (n-\tau)}.  \label{eq:chix}
\end{align}
Let $r\in \mathbb{Z}$ be a solution to $2\alpha v+\Delta =rN$. Due to (\ref{orthogonality}), there exists one unique $r$ when $v=0,1,\ldots, N-1$. Thus, (\ref{eq:chix}) is non-zero only when $v=(rN-\Delta)/2\alpha$. The modulus AF is obtained from (\ref{eq:conva}) with (\ref{eq:chig}) and (\ref{eq:chix}) as
\begin{align}
|\chi (\Delta,\tau)|=& \Bigg | \sum_{v=0}^{N-1} \delta[2\alpha v+\Delta \ (\mathrm{mod}\, N)] \notag \\
\times &\frac{1}{N}\sum_{n=0}^{N-1} x[n-\tau+v \ (\mathrm{mod}\, N)]  x^*[n]e^{j\frac{2\pi}{N}\Delta (n-\tau+v)} \Bigg | \notag \\
=& \Bigg |\frac{1}{N}\sum_{n=0}^{N-1} x\left [ n-\tau+\left ( \frac{rN-\Delta}{2\alpha}\right ) \ (\mathrm{mod}\, N) \right]\notag \\
 \times &x^*[n] e^{j\frac{2\pi}{N}\Delta n}\Bigg |.
\end{align}

\section{Proofs for Property 7 - 8}
\subsection{Proof for Property 7}
Case 1 $(u=2\alpha)$. Inserting $x_u[m]$ in (\ref{ccdt}) gives:
\begin{align}
\mathrm{PAPR}&=\! \! \! \! \max_{0 \le n \le N-1} \left | \frac{1}{\sqrt{N}} e^{-j\frac{2\pi}{N}(\alpha n^2 +\beta n + \gamma)} 
\sum_{m=0}^{N-1} e^{j\frac{2\pi}{N}(\alpha+2\alpha n +\beta)m} \right |^2 \notag \\
&=\! \! \! \! \max_{0 \le n \le N-1} \left |\frac{1}{\sqrt{N}} \sum_{m=0}^{N-1} e^{j\frac{2\pi}{N}(\alpha+2\alpha n +\beta)m} \right |^2 
\end{align}
If $\alpha\in\mathbb{Z}$, then it follows from (\ref{periodicity}) that $\beta\in\mathbb{Z}$, therefore $\alpha+2\alpha n +\beta\in\mathbb{Z}$, since $2\alpha \in \mathbb{Z}$. If $\alpha=p/2$ for any odd integer $p$, then there exists an odd integer $q$ such that $\beta=q/2$ and $\alpha+\beta=(p+q)/2\in\mathbb{Z}$, therefore $\alpha+2\alpha n +\beta\in\mathbb{Z}$. Thus
\begin{equation}
\label{eq:proof13}
\frac{1}{\sqrt{N}} \sum_{m=0}^{N-1} e^{j\frac{2\pi}{N}(\alpha+2\alpha n +\beta)m}=
\begin{cases}
\sqrt{N}, & \! \! \! \! \!\alpha+2\alpha n +\beta \equiv 0 \ (\mathrm{mod}\, N) \\
0, & \! \! \! \! \! \alpha+2\alpha n +\beta \nequiv 0 \ (\mathrm{mod}\, N) 
\end{cases}
\end{equation}
and we obtain $\max_{0 \le n \le N-1} |s[n]|^2=N$, i.e., $\mathrm{PAPR}=10 \log_{10} N$. 

Case 2 $(u\neq 2\alpha)$. Inserting $x_u[m]$ in (\ref{ccdt}) gives:
\begin{equation}
\mathrm{PAPR}=\max_{0 \le n \le N-1} \left | \sum_{m=0}^{N-1} e^{-j\frac{2\pi}{N}( (\alpha -\frac{u}{2})m^2 -(\frac{u}{2}+2\alpha n+\beta)m)} \right |^2 
\end{equation}
Let us define $a=\alpha -\frac{u}{2}$, $b=-(\frac{u}{2}+2\alpha n+\beta)$ and 
\begin{equation}
S= \left | \frac{1}{\sqrt{N}}\sum_{m=0}^{N-1} e^{-j\frac{2\pi}{N}( am^2 +bm)} \right |^2
\end{equation}
then $\mathrm{PAPR}=\max_{0 \le n \le N-1} S$ and
\begin{align}
S=&\frac{1}{N}\sum_{m=0}^{N-1} e^{-j\frac{2\pi}{N}( am^2 +bm)} \sum_{n=0}^{N-1} e^{j\frac{2\pi}{N}( an^2 +bn)} \notag \\
=&\frac{1}{N}\sum_{m=0}^{N-1} \sum_{n=0}^{N-1} e^{-j\frac{2\pi}{N}(m-n)(a(n+m)+b)} \notag \\
=&\frac{1}{N}\sum_{m=0}^{N-1} \sum_{n=0}^{N-1} e^{-j\frac{2\pi}{N}((m-n)2a n-(m-n)^2a+(m-n)b)}  \notag \\
=&\frac{1}{N}\sum_{t=0}^{N-1} \sum_{n=0}^{N-1} e^{-j\frac{2\pi}{N} (t2an-t^2a+tb)} \notag \\
=&\frac{1}{N}\sum_{t=0}^{N-1} \left ( \sum_{n=0}^{N-1}  e^{-j\frac{2\pi}{N}t2an} \right )e^{-j\frac{2\pi}{N} (-t^2a+tb)} \notag \\
=& \frac{1}{N}N
\end{align}
The last step follows since $t2a=t2\alpha-tu\in \mathbb{Z}$, $N$ is a prime, $\mathrm{gcd}(u,N)=1$ and thus $\mathrm{gcd}(t2a,N)=1$ and $t2\alpha \ (\mathrm{mod}\, N)\neq 0$. Hence, the inner sum is equal to $N$ when $t=0$. Therefore, 
 $\max_{0 \le n \le N-1} |s[n]|^2=1$ and $\mathrm{PAPR}=0$ dB.
 
 The variable substitution $t=m-n$  apply in the range $0\le t \le N-1$ since
 \begin{enumerate}[label=\roman*)]
 \item $e^{-j\frac{2\pi}{N}t2an}$
 \item $e^{-j\frac{2\pi}{N}(-t^2a+tb)}$
 \end{enumerate}
 have period of $N$. For i), it directly follows from that $t2a=t(2\alpha-u)$ is an integer. For ii), it can be shown as follows,
 \begin{align}
 e^{-j\frac{2\pi}{N}(-(t+N)^2a+b(t+N))}=&e^{-j\frac{2\pi}{N}(-t^2a+bt)}e^{-j2\pi(-Na-2ta+b)} \notag\\
 =&e^{-j\frac{2\pi}{N}(-t^2a+bt)}e^{j2\pi(N\alpha+\beta)} \notag\\
 &\times e^{j2\pi 2\alpha(t+n)}e^{-j2\pi\frac{u}{2}(N+2t-1)} \notag\\
 =&e^{-j\frac{2\pi}{N}(-t^2a+bt)} 
 \end{align}  
 where (\ref{periodicity}) and $2\alpha\in\mathbb{Z}$ are used, and since $\frac{u}{2}(N+2t-1)\in\mathbb{Z}$ for odd $N$.

\subsection{Proof for Property 8}
Inserting $x_k[m]$ in (\ref{ccdt}) gives
\begin{equation}
\mathrm{PAPR}=\max_{0 \le n \le N-1} \left | \sum_{m=0}^{N-1} e^{-j\frac{2\pi}{N}( \alpha m^2 -(k+2\alpha n+\beta)m)} \right |^2 
\end{equation}
Let us define $a=\alpha$, $b=-(k+2\alpha n+\beta)$ and perform the same steps as in Case 2 of the proof of Property 7.  Thus 
\begin{align}
S=&\frac{1}{N}\sum_{t=0}^{N-1} \left ( \sum_{n=0}^{N-1}  e^{-j\frac{2\pi}{N}t2an} \right )e^{-j\frac{2\pi}{N} (-t^2a+tb)}  
\end{align}
and it follows straightforwardly that 
$e^{-j2\pi(-Na-2ta+b)}=e^{-j2\pi(-N\alpha-\beta-2t\alpha-k-2\alpha n)}=1$ due to (\ref{periodicity}), $2\alpha\in\mathbb{Z}$, and $k$ and $t$ being integers. Therefore,  $e^{-j\frac{2\pi}{N}(-(t+N)^2a+(t+N)b)}=e^{-j\frac{2\pi}{N}(-t^2a+tb)}e^{-j2\pi(-Na-2ta+b)}$ has a period of $N$ and the same substitution $t=m-n$ in $S$ is applicable. Hence, $S=1$ and $\mathrm{PAPR}=0$ dB.

\section{Proofs for Property 9 - 14}
\subsection{Proof for Property 9}
By insertion of $x[m]$ in (\ref{ofdmamb}) and using (\ref{geosum})
\begin{align}
|\chi_\mathrm{OFDM} (\Delta=0,\tau)|&=\frac{1}{N} \left | \sum_{m=0}^{N-1}x[m]x^*[m] e^{-j\frac{2\pi}{N}\tau m} \right |\notag \\
&=
\begin{cases}
0, & \tau\neq 0\\
1, & \tau=0.
\end{cases}
\end{align}
\subsection{Proof for Property 10}
By insertion of $x_k[m]$ in (\ref{ofdmamb}) and using (\ref{geosum})
\begin{align}
|\chi_\mathrm{DFT-s-OFDM} (\Delta,\tau=0)|&=\frac{1}{N} \left | \sum_{m=0}^{N-1}x[m]x^*[m] e^{j\frac{2\pi}{N}\Delta m} \right | \notag \\
&=
\begin{cases}
0, & \Delta\neq 0\\
1, & \Delta=0.
\end{cases}
\end{align}
\subsection{Proof for Property 11}
By insertion of $x_k[m]$ in (\ref{ofdmamb}) and (\ref{geosum})
\begin{align}
|\chi_\mathrm{OFDM} (\Delta,\tau)|&=\frac{1}{N} \left | \sum_{m=0}^{N-1} e^{-j\frac{2\pi}{N}\tau m} \right | =
\begin{cases}
0, & \tau\neq 0\\
1,& \tau=0.
\end{cases}
\end{align}
\subsection{Proof for Property 12}
By insertion of $x_k[m]$ in (\ref{dftsofdmamb}) and using (\ref{geosum})
\begin{equation}
|\chi_\mathrm{DFT-s-OFDM} (\Delta=0,\tau)|=\frac{1}{N} \left | \sum_{m=0}^{N-1} e^{j\frac{2\pi}{N}\Delta m} \right |=
\begin{cases}
0, & \Delta\neq 0\\
1, & \Delta=0.
\end{cases}
\end{equation}

\subsection{Proof for Property 13}
By insertion of $x_u[m]$ in (\ref{ofdmamb}) and (\ref{geosum})
\begin{align}
|\chi_\mathrm{OFDM} (\Delta,\tau)|&=\frac{1}{N} \Bigg |\sum_{k=0}^{N-1}  \sum_{m=0}^{N-1} 
e^{j\frac{\pi}{N}uk(k+1)} e^{-j\frac{\pi}{N}um(m+1)}  \notag \\
 & \times e^{-j\frac{2\pi}{N}\tau m} \sum_{n=0}^{N-1}e^{j\frac{2\pi}{N}n(k-m+\Delta)}\Bigg | \notag \\
&\stackrel{(\mathrm{a})}{=}\frac{1}{N} \Bigg | \sum_{m=0}^{N-1} e^{-j\frac{2\pi}{N}m(-u(rN-\Delta)+\tau)} \Bigg | \notag\\
&\stackrel{(\mathrm{b})}{=}
\begin{cases}
0, & \tau+u\Delta \nequiv 0 \ (\mathrm{mod} \,N)\\
1,& \tau+u\Delta \equiv 0 \ (\mathrm{mod} \,N) 
\end{cases}
\end{align}
where (a) follows from that the inner sum is $\delta[k-m+\Delta \ (\mathrm{mod} \,N)]$ which gives $k=m-\Delta + rN$ for $r\in \mathbb{Z}$. Step (b) follows from that the sum is $\delta[-u(rN-\Delta)+\tau \ (\mathrm{mod} \,N)]=\delta[u\Delta+\tau \ (\mathrm{mod} \,N)]$.

\subsection{Proof for Property 14}
By insertion of $x_u[m]$ in (\ref{dftsofdmamb}) and (\ref{geosum})
\begin{align}
|\chi_\mathrm{DFT-s-OFDM} (\Delta=0,\tau)|=&\frac{1}{N} \left | \sum_{m=0}^{N-1} e^{j\frac{2\pi}{N}m(\Delta-\tau u)} \right |\\
=&
\begin{cases}
0, & \Delta-\tau u \nequiv 0 \ (\mathrm{mod} \,N) \\
1, & \Delta-\tau u \equiv 0 \ (\mathrm{mod} \,N).
\end{cases}
\end{align}

\end{document}